\documentclass[aps,prl,twocolumn,superscriptaddress,letterpaper]{revtex4-2}
\usepackage{graphicx}
\usepackage{bm}
\usepackage{amssymb}
\usepackage{color}
\usepackage{amsmath}
\usepackage[colorlinks=true,linkcolor=blue,anchorcolor=black,citecolor=blue,filecolor=black,menucolor=black,runcolor=black,urlcolor=blue]{hyperref}

\renewcommand{\figurename}{\textbf{Fig.}}
\renewcommand{\thefigure}{\textbf{\arabic{figure}}}

\begin{document}

\title{Stiefel-Whitney topological charges in a three-dimensional acoustic nodal-line crystal}

\author{Haoran Xue}
\affiliation{Division of Physics and Applied Physics, School of Physical and Mathematical Sciences, Nanyang Technological University,
Singapore 637371, Singapore}

\author{Z. Y. Chen}
\affiliation{National Laboratory of Solid State Microstructures and Department of Physics, Nanjing University, Nanjing 210093, China}

\author{Zheyu Cheng}
\affiliation{Division of Physics and Applied Physics, School of Physical and Mathematical Sciences, Nanyang Technological University,
	Singapore 637371, Singapore}

\author{J. X. Dai}
\affiliation{National Laboratory of Solid State Microstructures and Department of Physics, Nanjing University, Nanjing 210093, China}

\author{Yang Long}
\affiliation{Division of Physics and Applied Physics, School of Physical and Mathematical Sciences, Nanyang Technological University,
Singapore 637371, Singapore}

\author{Y. X. Zhao}
\email[]{zhaoyx@nju.edu.cn}
\affiliation{National Laboratory of Solid State Microstructures and Department of Physics, Nanjing University, Nanjing 210093, China}
\affiliation{Collaborative Innovation Center of Advanced Microstructures, Nanjing University, Nanjing 210093, China}

\author{Baile Zhang}
\email{blzhang@ntu.edu.sg}
\affiliation{Division of Physics and Applied Physics, School of Physical and Mathematical Sciences, Nanyang Technological University,
Singapore 637371, Singapore}
\affiliation{Centre for Disruptive Photonic Technologies, Nanyang Technological University, Singapore 637371, Singapore}

\maketitle

\textbf{Band topology of materials describes the extent Bloch wavefunctions are twisted in momentum space. Such descriptions rely on a set of topological invariants, generally referred to as topological charges, which form a characteristic class in the mathematical structure of fiber bundles associated with the Bloch wavefunctions. For example, the celebrated Chern number and its variants belong to the Chern class, characterizing topological charges for complex Bloch wavefunctions. Nevertheless, under the space-time inversion symmetry, Bloch wavefunctions can be purely real in the entire momentum space; consequently, their topological classification does not fall into the Chern class, but requires another characteristic class known as the Stiefel-Whitney class. Here, in a three-dimensional acoustic crystal, we demonstrate a topological nodal-line semimetal that is characterized by a doublet of topological charges, the first and second Stiefel-Whitney numbers, simultaneously. Such a doubly charged nodal line gives rise to a doubled bulk-boundary correspondence — while the first Stiefel–Whitney number induces ordinary drumhead states of the nodal line, the second Stiefel–Whitney number supports hinge Fermi arc states at odd inversion-related pairs of hinges. These results establish the Stiefel–Whitney topological charges as intrinsic topological invariants for topological materials, with their unique bulk-boundary correspondence beyond the conventional framework of topological band theory.}

Quantum mechanical wavefunctions are written in complex numbers, and so are the Bloch wavefunctions in crystals. These complex Bloch wavefunctions are twisted in momentum space to form band topology, following their mathematical structure of fiber bundles that is characterized by a set of topological invariants known as a characteristic class. A famous example of the topological invariant is the Chern number in the Chern class, which can be treated as a topological charge that induces topological boundary states, following the principle of bulk-boundary correspondence~\cite{thouless1982}. Such a correspondence from bulk to boundary is generally one to one, since different topological phases are incompatible and do not exist simultaneously to host different topological charges. Materials classified in the Chern class have been extensively explored for decades, leading to many discoveries such as the Chern insulators, time-reversal-invariant topological insulators, and Weyl semimetals~\cite{hasan2010, qi2011, haldane1988, qi2008, wan2011, armitage2018}. 

In the presence of symmetries, the properties of the Hamiltonian eigenspace can be significantly modified~\cite{chiu2016}. A prominent example is the spacetime inversion ($PT$) symmetry. In the field of non-Hermitian physics, $PT$ symmetry has played a central role as it can lead to real eigenenergies that are unexpected for a non-Hermitian Hamiltonian~\cite{bender1998}. In periodic Hermitian systems without spin-orbit coupling, while the eigenenergies are already real, the application of $PT$ symmetry is able to refine the Bloch wavefunctions from complex numbers to real numbers~\cite{Zhao2016,Zhao2017}. Accordingly, the Chern number must vanish in such a scenario, and the Chern class classification is no longer eligible. Instead, the Stiefel-Whitney (SW) class is responsible for the topological classification of the $PT$-symmetric systems with purely real eigenspaces~\cite{nakahara2018geometry}.

The SW class consists of two topological charges, the first and second SW numbers, classifying 1D and 2D $PT$-symmetric systems, respectively. A nontrivial first (second) SW number represents the obstruction of finding a global real basis of fiber bundles for Bloch wavefunctions in the 1D (2D) Brillouin zone~\cite{nakahara2018geometry}. This context is similar to the Chern number in the obstruction of finding a global complex basis for Bloch wavefunctions in the 2D Brillouin zone. While the first SW number is equivalent to the quantized Berry phase, the second SW number is unique to the SW class, being able to protect 2D higher-order topological insulators and 3D topological semimetals~\cite{Yang_2018prl, Yang_2019prx, Sigrist_2017prb, Sheng2019, wang2020, Sheng2022}, as counterparts of Chern insulators and Weyl semimetals protected by the Chern number. More intriguingly, recent theories suggest that nontrivial first and second SW numbers can co-exist in a single system~\cite{Yang_2018prl, Yang_2019prx, wang2020, Sheng2022}, leading to a doubled bulk-boundary correspondence — the same bulk can be doubly charged with two topological charges simultaneously, which give rise to two kinds of boundary states at different locations.

\begin{figure}
	\centering
	\includegraphics[width=\columnwidth]{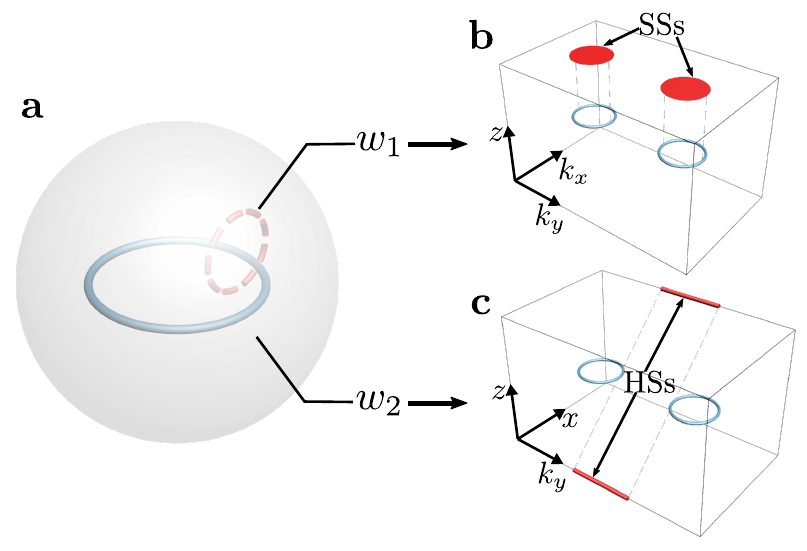}
	\caption{\textbf{Doubly charged nodal line.} \textbf{a}, Illustration of a nodal line with a doublet of topological charges $(w_1,w_2)$. $w_1$ and $w_2$ are defined on the chosen $S^1$ and $S^2$ surrounding the nodal line, respectively. \textbf{b}, Due to $w_1=1$, the surface states (SSs) form drumheads bounded by the projections of bulk nodal lines on the surface Brillouin zone. \textbf{c}, $w_2=1$ leads to a $PT$-related pair of hinge state (HS) Fermi arcs bounded by the projections of the bulk nodal lines on the hinge Brillouin zone.}
	\label{fig1}
\end{figure}

\begin{figure*}
	\centering
	\includegraphics[width=\textwidth]{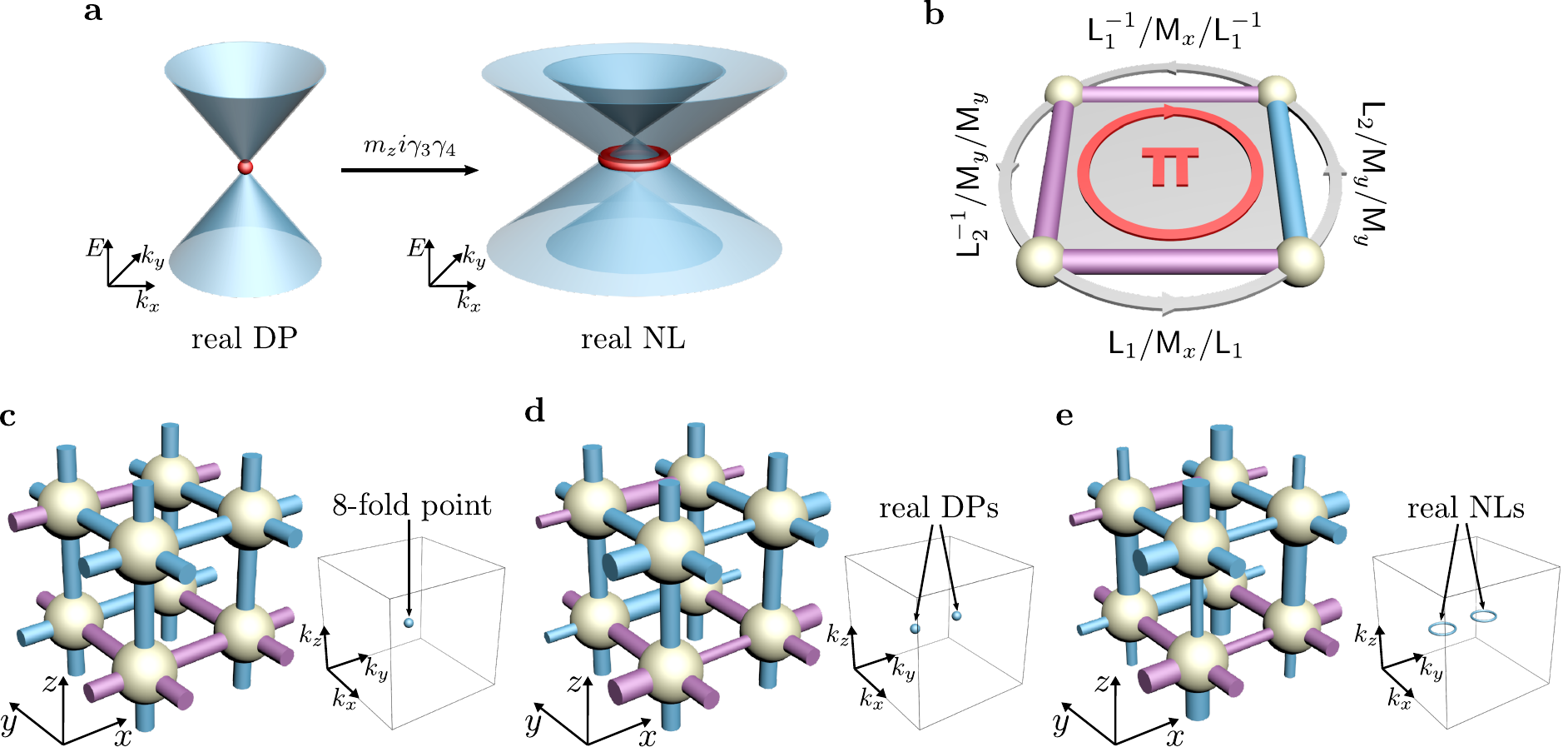}
	\caption{\textbf{Illustrations for the Dirac model, projective symmetry algebras and the tight-binding model.}  \textbf{a}, The fourfold degenerate real Dirac point is monotonically spread into a real nodal line by the partial mass term of $m_z\mathrm{i}\gamma_3\gamma_4$. \textbf{b}, Successively implementing operators sends a particle to circle the rectangular plaquette with flux $\pi$. Here, positive (negative) hopping amplitudes are colored in purple (blue). \textbf{c}, The undimerized lattice with flux $\pi$ through every rectangular plaquette. A topologically ``neutral'' eightfold degenerate crossing point resides at $K=(\pi,\pi,\pi)$ in the Brillouin zone. \textbf{d}, The dimerization $\mathfrak{D}_1$ is added along the $x$-direction, which is alternative along the $y$ direction and uniform along the $z$-direction. Hence, the eightfold degenerate crossing point is split into a pair of fourfold degenerate real Dirac points, each with topological charge $w_2=1$. \textbf{e}, The dimerization $\mathfrak{D}_2$ along the $z$ direction is further added, which alternates along both $x$ and $y$ directions. Then, each real Dirac point is spread into a real nodal line.  }
	\label{fig:Theory}
\end{figure*}

Here, in a three-dimensional (3D) acoustic crystal, we experimentally realize a nodal-line topological semimetal with a doublet of  SW topological charges as illustrated in Fig.~\ref{fig1}a, with $w_1$ and $w_2$ the first and second SW numbers, respectively (Supplementary Section 1). Such a nodal line can be named a real nodal line due to its purely real eigenspace. Because of the nontrivial $w_2$, these nodal lines appear in pairs (see Fig.~\ref{fig1}b and c)~\cite{Yang_2018prl, wang2020}, resembling the Nielsen–Ninomiya theorem of Weyl points in Weyl semimetals~\cite{nielsen1981absence}. The $1$D topological charge $w_1$ leads to the first-order drumhead surface states (SSs)~\cite{weng2015}, which also appear in the case of a conventional nodal line (see Fig.~\ref{fig1}b). However, the additional $2$D topological charge $w_2$, which is unique to the SW class, can give rise to odd $PT$-related pairs of hinge Fermi arcs. In our experiment, this is demonstrated by a sample of a long rectangular prism that hosts a single pair of $PT$-related gapless hinges (see Fig.~\ref{fig1}c and Fig.~\ref{fig4}). The novel distribution of hinge states (HSs) distinguishes this unconventional nodal-line semimetal from other existing second-order topological semimetals that host HSs on all four hinges~\cite{luo2021, wei2021, qiu2021}. 
This novelty is further experimentally confirmed on a sample with a more irregular but still $PT$-invariant geometry as shown in Fig.~\ref{fig5}. 

\textit{General idea.}
Let us start with introducing the minimal Dirac model for a nodal line with a doublet of topological charges $(w_1,w_2)$ (Supplementary section 2):
\begin{equation} \label{Real_Dirac}
	\mathcal{H}(\bm{k})=k_x\gamma_1+k_y\gamma_2+k_z\gamma_3+m_z\mathrm{i}\gamma_3\gamma_4.
\end{equation}
Here, $\gamma_a$ with $a=1,2,\cdots,5$ are the $4\times 4$ Hermitian Dirac matrices satisfying the Clifford algebra: $\{\gamma_a,\gamma_b\}=2\delta_{ab}1_4$.
Without loss of generality, we represent the $PT$ operator as $\mathcal{P}\mathcal{T}=\mathcal{K}$ with $\mathcal{K}$ the complex conjugation.  In model \eqref{Real_Dirac}, we have ordered the Dirac matrices so that $\gamma_i$ with $i=1,2,3$ are real and $\gamma_{4,5}$ are purely imaginary. A set of matrices representing $\gamma_a$ can be found in Methods. Hence, it is easy to check \eqref{Real_Dirac} is indeed a real Hamiltonian preserving $PT$ symmetry. Moreover, since $i\gamma_3\gamma_4$ anticommutes with $\gamma_3$ while commutes with $\gamma_{1,2}$, we may refer to $m_z\mathrm{i}\gamma_3\gamma_4$ as the partial mass term along the $k_z$ direction. The nodal line lies on the $k_x$-$k_y$ plane, and its radius increases monotonically as $m_z$ (see Fig.~\ref{fig:Theory}a). When $m_z=0$, the ring shrinks into a Dirac point named a real Dirac point due to its real eigenspace~\cite{Zhao2017}.

Inspired by this continuum model, we develop a lattice construction method briefly introduced as follows. First, we shall form an eightfold degenerate band crossing point in the momentum space. Although the crossing point is topologically ``neutral'', we then add appropriate symmetry-breaking dimerization patterns in order to split it into two fourfold degenerate real Dirac points, each with topological charge $w_2=1$. The last step is to further spread each point into a nodal line with certain appropriate dimerization patterns. 

\begin{figure*}
  \centering
  \includegraphics[width=\textwidth]{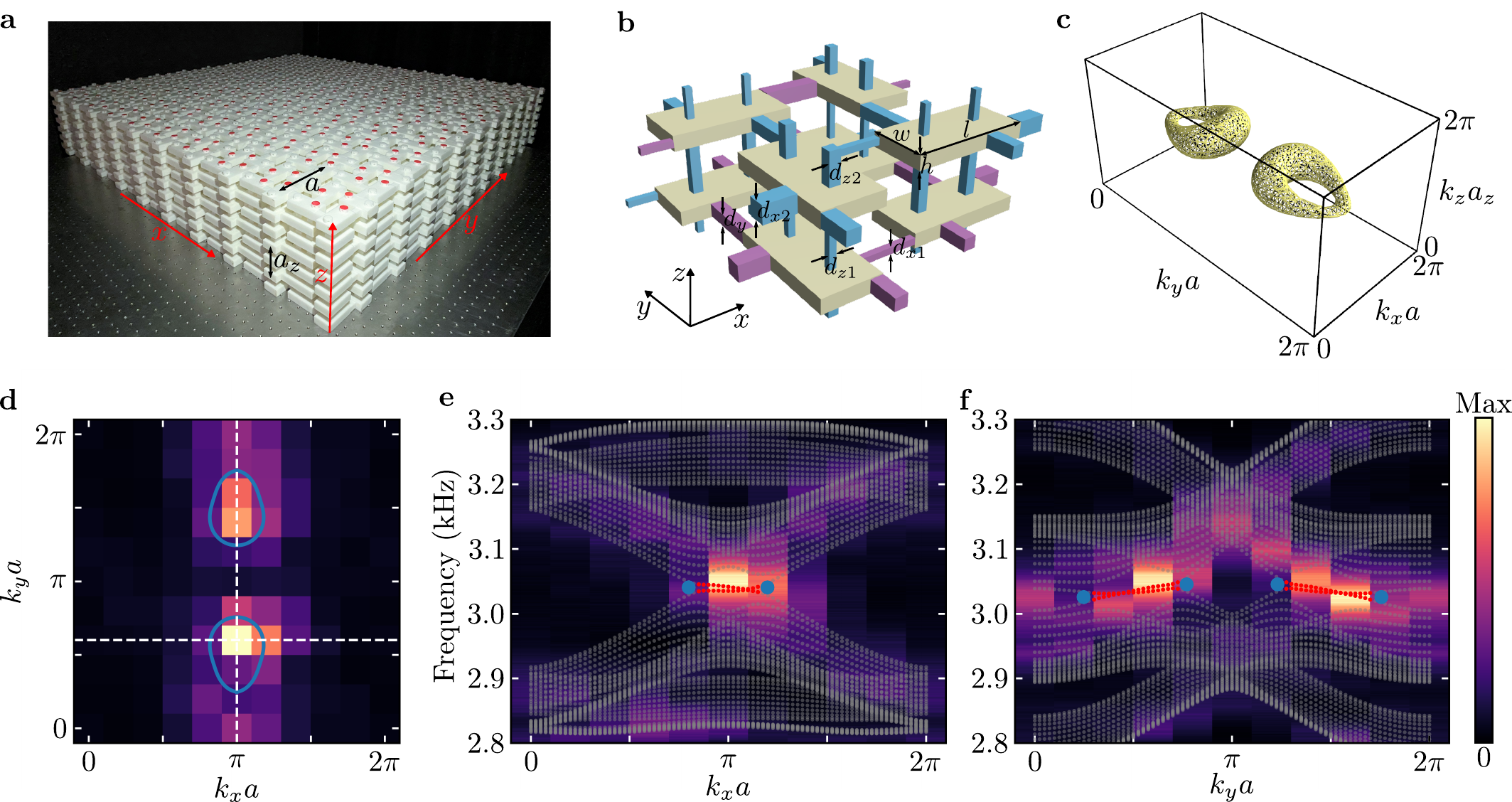}
  \caption{\textbf{Acoustic crystal design and the observation of drumhead surface states.} \textbf{a}, Experimental sample for measuring the SSs. The lattice constants in the $xy$ plane and along the $z$ direction are $a=140$ mm and $a_z=70$ mm, respectively. \textbf{b}, Unit cell of the acoustic crystal, with tubes enabling positive and negative couplings colored in purple and blue, respectively. The dimensions of the cuboid cavities are $l=56$ mm, $w=28$ mm and $h=7$ mm. The width parameters of the tubes are $d_{x1}=3.2$ mm, $d_{x2}=7.8$ mm, $d_y=6$ mm, $d_{z1}=3.2$ mm and $d_{z2}=4.8$ mm. The acoustic crystal is filled with air and surrounded by hard walls. \textbf{c}, Equi-frequency surface of the acoustic crystal in \textbf{b} at 3020 Hz (slightly below the frequency range of the nodal ring) calculated from full-wave simulations. \textbf{d}, Experimentally measured SS dispersion at 3045 Hz (inside the frequency range of the nodal ring). The solid blue curves denote the projections of the nodal rings. \textbf{e}(\textbf{f}), Experimentally measured SS dispersion along the horizontal (vertical) white dashed line in \textbf{d}. The grey and red dots represent simulated eigenfrequencies of the bulk and SSs, respectively. The blue dots denote the projections of the bulk nodal points.}
  \label{fig3}
\end{figure*}

\begin{figure*}
  \centering
  \includegraphics[width=\textwidth]{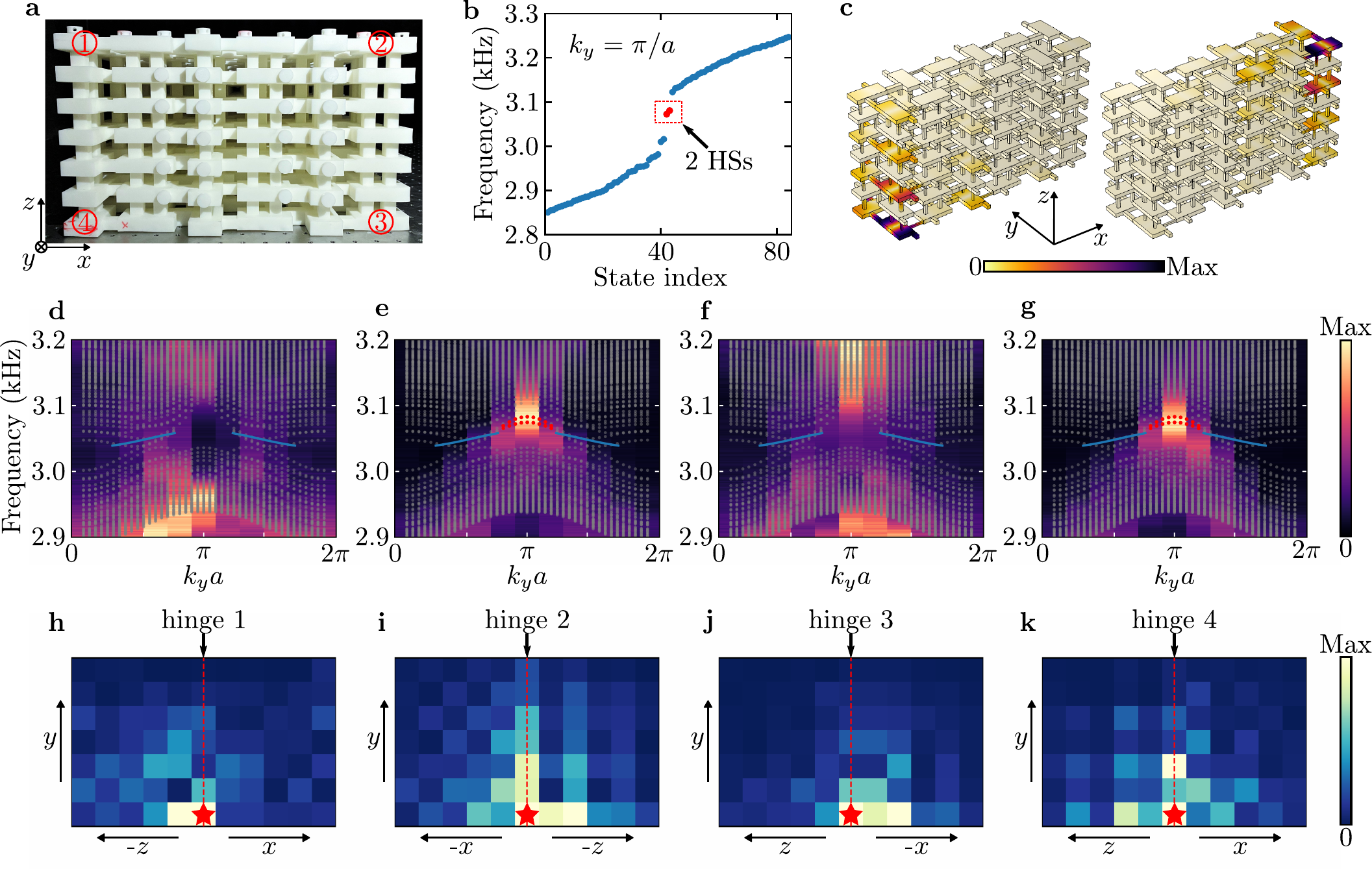}
  \caption{\textbf{Observation of $\bm{PT}$-related hinge states.} \textbf{a}, Experimental sample for measuring the HSs. The numbers \textcircled{1}--\textcircled{4} label the four $y$-directional hinges. \textbf{b}, Simulated eigenfrequencies for the sample shown in \textbf{a} at $k_y=\pi/a$. The blue dots represent the bulk and SSs, and the red dots indicate the HSs. \textbf{c}, Eigen profiles for the two states highlighted in red in \textbf{b}. The color indicates the amplitude of the acoustic pressure. \textbf{d}--\textbf{g}, Experimentally measured dispersions for hinge 1 (\textbf{d}), hinge 2 (\textbf{e}), hinge 3 (\textbf{f}) and hinge 4 (\textbf{g}). The grey dots represent simulated eigenfrequencies of the bulk and SSs, and the red dots indicate simulated hinge bands. The solid blue lines denote the projections of the nodal rings. \textbf{h}--\textbf{k}, Experimentally measured acoustic intensity distributions on two surfaces adjacent to hinge 1 (\textbf{h}), hinge 2 (\textbf{i}), hinge 3 (\textbf{j}) and hinge 4 (\textbf{k}). The red star indicates the position of the speaker and the red dashed line highlights the position of the hinge. The operating frequencies of the speaker are chosen as: 3078 Hz (hinge 1 and hinge 3), 3076 Hz (hinge 2) and 3080 Hz (hinge 4), which are around the eigenfrequencies of the HSs.}
  \label{fig4}
\end{figure*} 

To achieve a nodal point with high degeneracy, we utilize projective symmetries, which stem from the gauge fluxes on the lattice and can lead to high dimensional irreducible representations~\cite{Wen-PSG,Zhao_2020prb,Shao_2021prl}. Moreover, the $\mathbb{Z}_2$ lattice gauge fields are highly engineerable in artificial lattices (i.e., the sign of each real hopping amplitude can be flexibly tuned to be $+$ or $-$ ), as demonstrated in recently realized projectively symmetry-protected topological phases in acoustic crystals~\cite{xue2022,li2022}.

\textit{Model construction.}
As aforementioned, to construct a realizable lattice model, we have recourse to the projective symmetry algebra. We consider a $3$D rectangular lattice with the nearest-neighbor hoppings, which has flux $\pi$ for each rectangular plaquette along any direction. The flux pattern can be described by numerous configurations of signs of hopping amplitudes, and the one we choose is given in Fig.~\ref{fig:Theory}c-e. Because of the gauge fluxes, the unit translation operators $\mathsf{L}_i$ with $i=x,y,z$, which previously mutually commute, become pairwisely anti-commuting, i.e., $\{\mathsf{L}_i,\mathsf{L}_j\}=0$ for $i\ne j$, which constitute the projective symmetry algebra of translations. The anti-commutation relation manifests the Aharonov–Bohm effect, since the equivalent form $\mathsf{L}_j^{-1}\mathsf{L}_i^{-1}\mathsf{L}_j\mathsf{L}_i=-1$ corresponds to that a particle accumulates a phase factor $e^{i\pi}=-1$ after circling the plaquette spanned by $\mathsf{L}_i$ and $\mathsf{L}_j$, as illustrated in Fig.~\ref{fig:Theory}b. Similar analysis shows the projective algebraic relations $\{\mathsf{M}_i,\mathsf{M}_j\}=2\delta_{ij}$ for mirror reflections $\mathsf{M}_i$, and those between $\mathsf{M}_i$ and $\mathsf{L}_j$: $\mathsf{M}_i\mathsf{L}_i\mathsf{M}_i=\mathsf{L}_i^{-1}$ and $\{\mathsf{M}_i,\mathsf{L}_j\}=0$ for $i\ne j$ (see Fig.~\ref{fig:Theory}b). Here, $\mathsf{M}_i$ reverses the $i$th coordinate with the center of the unit cell as the coordinate origin. We now turn to the Brillouin zone, and denote the representations of $\mathsf{L}_i$ and $\mathsf{M}_i$ as $\mathcal{L}_i^{\bm{k}}$ and $\mathcal{M}_i$, respectively. Note that $\mathcal{L}_i^{\bm{k}}$ depend on $\bm{k}$, while $\mathcal{M}_i$ are independent of $\bm{k}$. Specializing at the point $K=(\pi,\pi,\pi)$, their projective algebraic relations are given by
\begin{equation}\label{M_PSA}
	-\{\mathcal{L}_i^{K},\mathcal{L}_j^{K}\}= \{\mathcal{M}_i,\mathcal{M}_j\}=2\delta_{ij}, ~~~\{\mathcal{M}_i,\mathcal{L}^K_j\}=0.
\end{equation}
Since time reversal $T$ is preserved at $K$, we further consider its representation $\mathcal{T}$, which commutes with all $\mathcal{L}_i^{K}$ and $\mathcal{M}_i$.  The projective symmetry algebra generated by $\mathcal{L}_i^{K}$, $\mathcal{M}_i$ and $\mathcal{T}$ is equivalent to the real Clifford algebra $C^{3,3}\otimes C^{1,1}\cong C^{4,4}\cong \mathbb{R}(2^{d+1})$, which has a unique complex irreducible representation with dimension $2^{3}$. That is, there is the desired eightfold degenerate crossing point at $K$ protected by the projective symmetry algebra \eqref{M_PSA}. Since our unit cell consists of $8$ sites, all bands are enforced to cross at $K$ to represent \eqref{M_PSA}.

\begin{figure*}
  \centering
  \includegraphics[width=0.7\textwidth]{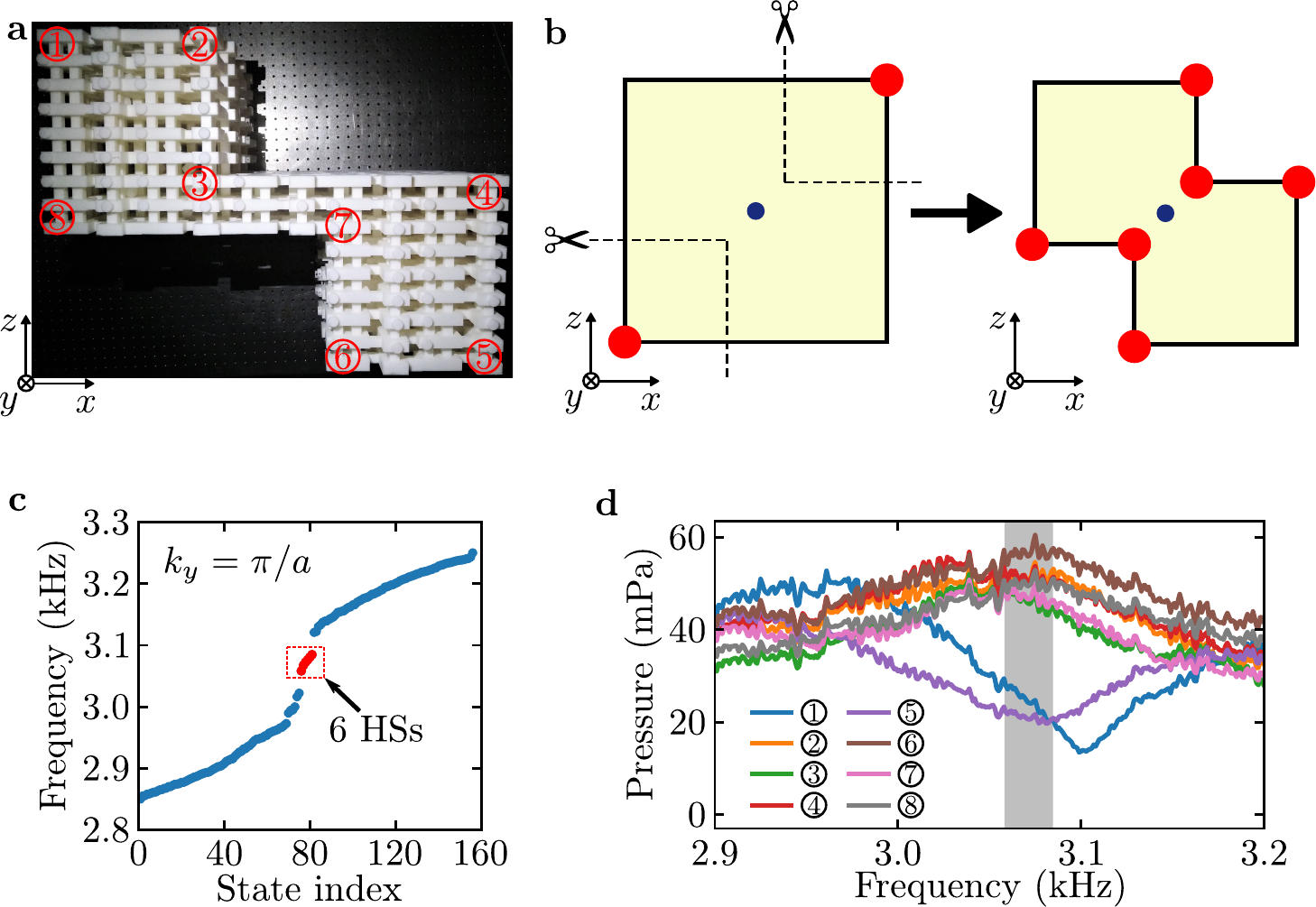}
  \caption{\textbf{Increased number of hinge states.} \textbf{a} Experimental sample for demonstrating the odd number of pairs of $PT$-related HSs. The numbers \textcircled{1}--\textcircled{8} label the eight $y$-directional hinges. \textbf{b}, Illustration of the process of obtaining an irregularly shaped sample supporting three pairs of HSs by cutting two corners of a rectangle sample with one pair of HSs. Here the red circles represent the HSs and the blue circles denote the inversion centers of the samples. \textbf{c}, Simulated eigenfrequencies for the sample shown in \textbf{a} at $k_y=\pi/a$. The blue dots represent the bulk and SSs, and the red dots indicate the HSs. \textbf{d}, Experimentally measured transmission spectra at the eight hinges. The grey region indicates the frequency range of the HSs.}
  \label{fig5}
\end{figure*} 

To have a nontrivial SW class, we consider the inversion symmetry $\mathsf{P}$ centered at the upper-right $y$-bond of the unit cell in Fig.~\ref{fig:Theory}c, i.e., $\mathsf{P}=\mathsf{L}_x\mathsf{L}_z\mathsf{M}_x\mathsf{M}_y\mathsf{M}_z$. Following the aforementioned method, we should proceed to add $\mathsf{P}T$-invariant dimerization patterns that break some $\mathsf{L}_i$ and $\mathsf{M}_j$ to realize the real nodal lines. Let us consider two dimerization patterns in order. The first pattern $\mathfrak{D}_1$ is the dimerization along the $x$ direction, which alternates along the $y$ direction but is invariant along the $z$ direction as illustrated in Fig.~\ref{fig:Theory}d. Adding $\mathfrak{D}_1$ splits the eightfold degenerate topologically ``neutral'' crossing point into two fourfold degenerate real Dirac points charged by $w_2$, each of which is modeled by \eqref{Real_Dirac} with $m_z=0$. We then introduce the second dimerization pattern $\mathfrak{D}_2$, which gives rise to the partial mass term $m_z \mathrm{i}\gamma_3\gamma_4$ of \eqref{Real_Dirac} and therefore can spread each real Dirac point into a nodal line. $\mathfrak{D}_2$ is designed as the dimerization along the $z$-direction, which alternates along both the $x$ and $y$ directions, as illustrated in Fig.~\ref{fig:Theory}e. Each of the nodal lines, as shown in Supplementary section 3, carries a nontrivial doublet of topological charges $(w_1,w_2)$ as expected.

This tight-binding model with $\pi$ fluxes on a simple rectangular lattice can be implemented using the coupled acoustic cavity structure \cite{ma2019, xue2022b, xue2019, ni2019, li2018, xue2020, ni2020, qi2020, xue2022, li2022}. Our designed acoustic crystal is shown in Fig.~\ref{fig3}a, with each site in the tight-binding model implemented by a cuboid cavity supporting a dipolar resonance at around 3100 Hz. The coupling between two cavities is enabled by a tube with a square cross-section, with the coupling sign determined by the position of the tube and the coupling amplitude controlled by the width of the tube. Thus, by carefully engineering the coupling tubes, the required gauge fluxes and coupling dimerizations can be realized (see Fig.~\ref{fig3}b). The whole structure is filled with air and surrounded by photosensitive resins that act as sound rigid walls (see Methods). Figure~\ref{fig3}c shows the simulated equi-frequency surface of the acoustic crystal at 3020 Hz (slightly below the minimum frequency of the nodal line), which reveals the existence of two nodal lines forming two rings in the bulk dispersion and suggests the validity of this acoustic design (see Supplementary section 4 for more details on the dispersion).

\textit{Drumhead surface states.} We first demonstrate the existence of conventional drumhead SSs due to the nontrivial first-order topology induced by $\omega_1$. To this end, we scan the acoustic fields on the top surface excited by a speaker placed at the surface center (see Methods and Extended Data Fig.~\ref{fige1}). Figure~\ref{fig3}d shows the corresponding Fourier intensity at 3045 Hz, where the hot spots occur at positions inside the projections of the nodal rings (denoted by the blue lines), consistent with the feature of the drumhead SSs. To further confirm the existence of the drumhead SSs, we also plot in Fig.~\ref{fig3}e, f the frequency-resolved Fourier spectra along the $k_x$ and $k_y$ momenta, respectively (indicated by the two white dashed lines in Fig.~\ref{fig3}d). As can be seen, the measured SSs connect the projections of the two bulk nodal points (indicated by the blue dots) from the same nodal ring,  matching well with the simulated SS dispersion (indicated by the red dots).

\textit{PT-related hinge states.}
Next, we study the unconventional higher-order topology in this acoustic crystal induced by the nontrivial second SW number $w_2$. Let us first consider a sample with a simple rectangular geometry as shown in Fig.~\ref{fig4}a. This sample consists of six (seven) cavities in the $x$ ($z$) direction, thus preserving the required $PT$ symmetry. By imposing periodic boundary condition along the $y$ direction, we can numerically compute the dispersion for $y$-directional hinges. The results for $k_y=\pi/a$ are shown in Fig.~\ref{fig4}b and the results for all $k_y$ are given as colored dots in Fig.~\ref{fig4}d-g. As can be seen, there are two bands connecting the projections of the nodal rings. The eigen profiles of these two states are given in Fig.~\ref{fig4}c, showing they are indeed the hinge Fermi arcs states. Notably, the HSs only exist on two of the four hinges related by the $PT$ symmetry. Interestingly, the locations of the HSs can be transferred from the two off-diagonal hinges to the two diagonal ones by simply reversing the dimerization along $z$ (i.e., swap the values of $d_{z1}$ and $d_{z2}$; see Extended Data Fig.~\ref{fige2}). 

To probe the hinge Fermi arcs, we place a speaker at the hinge and scan the acoustic field along the hinge. This experiment is repeated for all four hinges (see the labelings of the hinges in Fig.~\ref{fig4}a) and the measured dispersions are plotted in Fig.~\ref{fig4}d-g. For hinge 2 (Fig.~\ref{fig4}e) and hinge 4 (Fig.~\ref{fig4}g), the measured dispersions match with the simulated hinge Fermi arcs (red dots), suggesting the existence of HSs on these two hinges. In contrast, the excited states are bulk states when the speaker is placed at hinge 1 (Fig.~\ref{fig4}d) or hinge 3 (Fig.~\ref{fig4}f). These results demonstrate the off-diagonal distribution (i.e., only at hinge 2 and hinge 4) of the HSs. In addition to the momentum space results, we also conduct real space measurements to reveal the HSs. For each hinge, a speaker operating at the HS's frequency is placed at one end and the acoustic field on the two adjacent surfaces is measured. As shown in  Fig.~\ref{fig4}h-k, the measured acoustic intensity distributions exhibit a hinge localization profile only for hinge 2 (Fig.~\ref{fig4}i) and hinge 4 (Fig.~\ref{fig4}k), which is consistent with the information from the simulated and measured dispersions. Note that here the HSs do not show a clear propagation along the hinge due to their small group velocity and the system's loss. Nevertheless, the strong field enhancement along the hinge is a clear signature of the HSs. 

A fascinating aspect of this acoustic crystal is that the protecting symmetry, i.e., the $PT$ symmetry, can be easily preserved in various geometries, not limited to regular ones like the square and rectangular geometries. Furthermore, under different geometries with the same bulk invariant $w_2$, the forms of the HSs (e.g., the locations and number of HSs) can also be different (see Supplementary section 5). This allows us to steer the HSs even without changing the system parameters. To demonstrate this nice property, we construct a sample with an irregular shape in the $xz$ plane, as shown in Fig.~\ref{fig5}a. One can imagine a cutting procedure illustrated in Fig.~\ref{fig5}b, where the two off-diagonal hinges of a rectangle-shaped sample are cut to get this irregularly-shaped sample. During such a process, the $PT$ symmetry remains intact while the pair of hinges that support the HSs are removed. Interestingly, the generated new sample host three, instead of one, pair of $PT$-related HSs (see Fig.~\ref{fig5}c). To demonstrate this phenomenon, we measure the transmission spectra at the eight hinges of this sample. As shown in Fig.~\ref{fig5}d, in the frequency range of the HSs, the signals measured at hinge 1 and hinge 5 are much lower than the signals measured at other six hinges. This indicates there are no HSs at hinge 1 and hinge 5, consistent with the simulation (see Extended Data Fig.~\ref{fige3}). The existence/absence of the HS at each hinge is also confirmed by real-sapce field measurements, as given in Extended Data Fig.~\ref{fige5}. 

\textit{Conclusion.}
In summary,  we have demonstrated an acoustic real nodal-line crystal in the nontrivial SW class, hosting ordinary drumhead SSs and unconventional $PT$-related HSs featuring exotic properties. Our study opens a new route to experimental studies on band topology constructed from real fiber bundles that were hardly explored previously. While our demonstration is in acoustics, the proposed idea can also be generalized to other classical wave systems with $PT$ symmetry, including photonic, mechanical and circuit systems \cite{serra2018, peterson2018, noh2018, imhof2018}. Besides, our results reveal the power of projective symmetries in realizing novel topological phases of matter, which could inspire more topological designs using artificial structures with gauge flux. On the practical level, the $PT$-related HSs, with superior tunability in their number and configuration compared to previous higher-order topological states, can provide robust and reconfigurable control over sound and other classical waves.

\bigskip
\noindent{\large{\bf{Methods}}}\\
\noindent{\textbf{Concrete form of the model and its symmetry operators}}\\
Since each unit cell consists of $2\times 2\times 2$ sites, we assign three sets of the standard Pauli matrices $\sigma$, $\tau$ and $\rho$ for the three dimensions $x$, $y$ and $z$, respectively. Moreover, we introduce the seven Hermitian $8\times 8$ Dirac matrices as
\begin{equation*}
	\begin{split}
		& \Gamma_1 = \rho_3 \otimes \tau_3 \otimes \sigma_2, \quad	\Gamma_2 = \rho_3 \otimes \tau_3 \otimes \sigma_1,   \\
		& \Gamma_3 = \rho_3 \otimes \tau_2 \otimes \sigma_0, \quad 	\Gamma_4 = \rho_3 \otimes \tau_1 \otimes \sigma_0,   \\
		& \Gamma_5 = \rho_2 \otimes \tau_0 \otimes \sigma_0, \quad 	\Gamma_6 = \rho_1 \otimes \tau_0 \otimes \sigma_0,   \\
		& \Gamma_7 =  \rho_3 \otimes \tau_3 \otimes \sigma_3.
	\end{split}
\end{equation*}
The Dirac matrices $\Gamma^\alpha$ with $\alpha=1,2\cdots,7$ satisfy $\{\Gamma^\alpha,\Gamma^\beta\}=2\delta^{\alpha\beta}$. With the Dirac matrices, the tight-binding Hamiltonian in momentum space is given by
\begin{equation} \label{Lattice_Model}
		\begin{split}
			H(\bm{k})= & \sum_{i,a}f_{i,a}(k_i)\Gamma_{2i-a+1}\\
			&+g_{x,1}(k_x)\mathrm{i}\Gamma_{2}\Gamma_{3} \Gamma_{4} +g_{x,2}(k_x)\mathrm{i}\Gamma_{1}\Gamma_{3}\Gamma_{4}\\
			&+g_{z,1}(k_z)\mathrm{i}\Gamma_{5}\Gamma_{7}+g_{z,2}(k_z)\mathrm{i}\Gamma_{6}\Gamma_{7}
		\end{split}	
\end{equation}
Let $t$ denote the hopping magnitude along the undimerized $y$ direction, and $J_{1}^{x,z}$ and $J_2^{x,z}$ be the two hopping magnitudes along the two dimerized directions $x$ and $z$, respectively. Then, the dimerization strengths are measured by $J^{x,z}_{-}/J_{+}^{x,z}$ with $J^{x,z}_{\pm}=(J_1^{x,z}\pm J_2^{x,z})/2$. The coefficient functions are given by $f_{x,1}(k_x)=-J_+^x(1+\cos k_x)$, $f_{x,2}=J_+^x\sin k_x$, $f_{y,1}(k_y)=-t(1+\cos k_y)$, $f_{y,2}(k_y)=t\sin k_y$, $f_{z,1}(k_z)=J_+^z(1+\cos k_z)$, $f_{z,2}(k_z)=-J^z_{+}\sin k_z$, and $g_{x,1}(k_x)=-J_{-}^x(1-\cos k_x)$, $g_{x,2}(k_x)=-J^x_{-}\sin k_x$,  $g_{z,1}(k_z)=-J_{-}^z(1-\cos k_z)$, $g_{z,2}(k_z)=J^z_{-}\sin k_z$. 

The momentum-space symmetry operators for $\mathsf{M}_i$ are given by
\begin{equation*}
	\begin{split}
		& \mathcal{M}_x = \rho_0 \otimes \tau_0 \otimes \sigma_1 \ \hat{m}_x =  -\mathrm{i} \Gamma_1 \Gamma_7 \hat{m}_x ,\\
		& \mathcal{M}_y = \rho_0 \otimes \tau_1 \otimes \sigma_3 \ \hat{m}_y = -\mathrm{i} \Gamma_3 \Gamma_7 \hat{m}_y,\\
		&	\mathcal{M}_z = \rho_1 \otimes \tau_3 \otimes \sigma_3 \ \hat{m}_z = -\mathrm{i} \Gamma_5 \Gamma_7 \hat{m}_z,\\
			\end{split}
\end{equation*}
where each $\hat{m}_i$ is the operator sending $k_i$ to $-k_i$ with $i=x,y,z$. It is easy to check the desired projective algebraic relations: $\{\mathcal{M}_i,\mathcal{M}_j\}=2\delta_{ij}$. The momentum-space operators for the translations $\mathsf{L}_i$ are given by
\begin{equation*}
	\begin{split}
		& \mathcal{L}_x = \rho_0 \otimes \tau_0 \otimes \begin{bmatrix}
			0 & e^{\mathrm{i}  k_x} \\
			1 & 0
		\end{bmatrix} ,  \\
		& \mathcal{L}_y = \rho_0  \otimes \begin{bmatrix}
			0 & e^{\mathrm{i}  k_y} \\
			1 & 0
		\end{bmatrix} \otimes \sigma_3, \\
		& \mathcal{L}_z  =  \begin{bmatrix}
			0 & e^{\mathrm{i}  k_z} \\
			1 & 0
		\end{bmatrix} \otimes \tau_3 \otimes \sigma_3.
	\end{split}
\end{equation*} 
We see that $\mathcal{L}_i^2=e^{\mathrm{i}k_i}$ for $i=x,y,z$, and $\{\mathcal{L}_i,\mathcal{L}_j\}=0$ if $i\ne j$. One can furthermore to check the projective algebraic relations between $\mathcal{L}_i$ and $\mathcal{M}_j$: $\{\mathcal{M}_i,\mathcal{L}_j\}=0$ if $i\ne j$, and $\mathcal{M}_i\mathcal{L}_i\mathcal{M}_i=-\mathcal{L}_i^\dagger$ for $i,j=x,y,z$.

It is easy to check that all symmetry operators $\mathcal{L}_i$ and $\mathcal{M}_j$ commute with time-reversal operator $\mathcal{T}=\mathcal{K}\hat{I}$ with $\hat{I}$ the inversion of $\bm{k}$ and $\mathcal{K}$ the complex conjugation. Specifically at $K=(\pi,\pi,\pi)$, we see
\begin{equation*}
	\begin{split}
			\mathcal{L}_x^K &=\rho_0\otimes\tau_0\otimes (-\mathrm{i}\sigma_2),\\
			 \mathcal{L}_y^K &=\rho_0\otimes (-\mathrm{i}\tau_2) \otimes \sigma_3, \\
			 \mathcal{L}_z^K &=(-\mathrm{i}\rho_2)\otimes \tau_3 \otimes \sigma_3.
	\end{split}
\end{equation*}
Together with operators $\mathcal{M}_i$ above, we can verify that the projective algebraic relations \eqref{M_PSA} at $K$ are indeed satisfied.

In the absence of dimerization, i.e., $J^x_-=J^z_-=0$, it is straightforward to check that all $\mathcal{M}_i$ and $\mathcal{L}_j$ commute with the Hamiltonian \eqref{Lattice_Model}. After the dimerization patterns $\mathfrak{D}_1$ and $\mathfrak{D}_2$ are introduced, all  $\mathsf{L}_i$ and $\mathsf{M}_j$ are broken, and therefore $\mathcal{M}_i$ and $\mathcal{L}_j$ do not commute with \eqref{Lattice_Model} any more. Nevertheless, the off-centered inversion symmetry $\mathsf{P} = \mathsf{L}_x \mathsf{L}_z \mathsf{M}_x  \mathsf{M}_y \mathsf{M}_z $ is preserved, and the momentum-space operator $\mathcal{P}$ for $\mathsf{P}$ can be derived as a product of the corresponding operators presented above. Then, the $\mathsf{P}T$ symmetry operator $\mathcal{P} \mathcal{T}$ is given by
	\begin{eqnarray}
		&&	\mathcal{P} \mathcal{T} =  \begin{bmatrix}
			e^{ \mathrm{i}  k_z} & 0 \\
			0 & 1
		\end{bmatrix} \otimes \tau_1 \otimes \begin{bmatrix}
			e^{ \mathrm{i}  k_x} & 0 \\
			0 & -1
		\end{bmatrix} \mathcal{K}.
	\end{eqnarray}
It is straightforward to check $\mathcal{P}\mathcal{T}$ commutes with the Hamiltonian \eqref{Lattice_Model} even with nonzero $J^x_{-}$ and $J^z_{-}$.

With small dimerizations $\mathfrak{D}_1$ and $\mathfrak{D}_2$, the low-energy effective model can be derived in the vicinity of each nodal line. Each low-energy effective model can be cast into the form of \eqref{Real_Dirac}, namely the real Dirac model with a ``partial mass term'' along the $z$ direction, by appropriately choosing the basis of four low-energy modes. The $4\times 4$ Hermitian Dirac matrices in \eqref{Real_Dirac} can be chosen as
\begin{equation*}
	\begin{split}
		\gamma_1=\sigma_1\otimes \tau_0,\quad \gamma_2=&\sigma_2\otimes\tau_2,\quad \gamma_3=\sigma_3\otimes\tau_0\\
		\gamma_4=\sigma_2\otimes\tau_3, &\quad  \gamma_5=\sigma_2\otimes\tau_1.
	\end{split}	
\end{equation*} 
Here, $\sigma$ and $\tau$ are two sets of the standard Pauli matrices, which have no direct relation with those used to define $\Gamma^\alpha$.

\noindent{\textbf{Full-wave simulation}}

All numerical simulations of the acoustic model are performed in the commercial software Comsol Multiphysics, pressure acoustics module. The software solves the Helmholtz equation with the finite element method. In the simulations, periodic boundary conditions with Bloch phase shifts are assigned to the periodic boundaries, while other boundaries are set as sound rigid boundaries due to the large impedance mismatch between the printing materials and air. The sound speed and density of air are set to be 347.2 m/s  and 1.16 $\text{kg/m}^3$, respectively. The geometrical parameters of the acoustic model are given in the caption of Fig.~\ref{fig3}.

To get the equi-frequency contour of the bulk bands (Fig.~\ref{fig3}c), we compute all the eight bulk bands in the area $0.5\pi/a<k_x<1.5\pi/a$, $0<k_y<\pi/a$ and $0.5\pi/a_z<k_z<1.5\pi/a_z$, with 30 computing points in each momentum direction. The contour at the other side of the Brillouin zone is obtained through the time-reversal operation. In the simulation of surface dispersion (Fig.~\ref{fig3}d-f), we adopt an acoustic supercell with periodic boundary condition along the $x$ and $y$ directions and 21 cavities along the $z$ direction. In the simulations of the hinge dispersions (Figs.~\ref{fig4}b-g and \ref{fig5}c), the simulated geometries in the $xz$ plane are the same as the experimental samples (see Figs.~\ref{fig4}a and \ref{fig5}a), with periodic boundary conditions imposed for the $y$ direction.

\noindent{\textbf{Sample design and fabrication}}

To implement the sound rigid walls that surround the air cavities and tubes, we design hard walls with a thickness of 5 mm to cover the whole sample. These walls are thick enough to provide the rigid wall condition. In order to excite and measure the sound signals, we drill two small holes (with radii of 5 mm) on the boundary cavities. These holes are covered with size-matched plugs when they are not in use.

The samples are fabricated through the stereolithography apparatus technique, with a fabrication resolution of around 0.1 mm. The dimensions of the three samples (i.e., the samples shown in Figs.~\ref{fig3}a, \ref{fig4}a and \ref{fig5}a) are around 1.5 m $\times$ 1.5 m $\times$ 0.2 m, 0.4 m $\times$ 1.5 m $\times$ 0.2 m and 0.7 m $\times$ 0.5 m $\times$ 0.5 m, respectively. Due to their large sizes, these samples are fabricated as separate parts and then assembled together.

\noindent{\textbf{Experimental measurement}}

All experiments are conducted using the same scheme, as illustrated in Extended Data Fig.~\ref{fige1}. The sound signal is launched by a speaker (Tymphany PMT-40N25AL01-04) placed on the surface (for measuring the SSs) or the hinge (for measuring the HSs). The speaker generates a broadband sound signal from 2000 Hz to 4000 Hz, which covers the frequency range of our interested bands. Two microphones (Br\"uel \& Kjær Type 4182) are used to detect the amplitude and phase of sound in the sample. One of the microphones is placed at the position of the source, serving as a reference probe. To ensure the accuracy of the frequency-resolved spectra, we have also checked that there are no resonances in the spectrum of the source (see Extended Data Fig.~\ref{fige4}). The other microphone scan field distributions in the targeted areas in the sample. The measured signal is processed by an analyzer (Br\"uel \& Kjær 3160-A-022 module) to obtain the experimental data with the amplitude and phase of sound at each measured point for the frequencies ranging from 0 Hz to 6400 Hz (the frequency resolution is 1 Hz). 

\bigskip
\noindent{\large{\bf{Data availability}}}

\noindent The experimental data are available in the data repository for Nanyang Technological University at this link (URL to be inserted upon publication). Other data that support the findings of this study are available from the corresponding authors upon reasonable request.

\bigskip
\noindent{\large{\bf{Acknowledgements}}}

\noindent H.X., Z.C., Y.L. and B.Z. are supported by the Singapore Ministry of Education Academic Research
Fund Tier 2 (Grant No.~MOE2019-T2-2-085) and Singapore National Research Foundation Competitive Research Program (Grant No.~NRF-CRP23-2019-0007). Z.Y.C., J.X.D. and Y.X.Z. acknowledge support from the National Natural Science Foundation of China (Grants No.~12161160315 and No.~12174181).

\bigskip
\noindent{\large{\bf{Author contributions}}}

\noindent H.X., Y.X.Z. and B.Z. conceived the idea. Z.Y.C., J.X.D. and Y.X.Z. developed the theory and the lattice model. H.X. constructed the acoustic model and performed numerical calculations. H.X., Z.C. and Y.L. conducted the experiments. H.X., Y.X.Z. and B.Z. wrote the manuscript with input from all authors. Y.X.Z. and B.Z. supervised the project. 

\bigskip
\noindent{\large{\bf{Competing interests}}}

\noindent The authors declare no competing interests.

\setcounter{figure}{0}
\renewcommand{\figurename}{\textbf{Extended Data Fig.}}
\renewcommand{\thefigure}{\textbf{\arabic{figure}}}
\begin{figure*}
  \centering
  \includegraphics[width=0.6\textwidth]{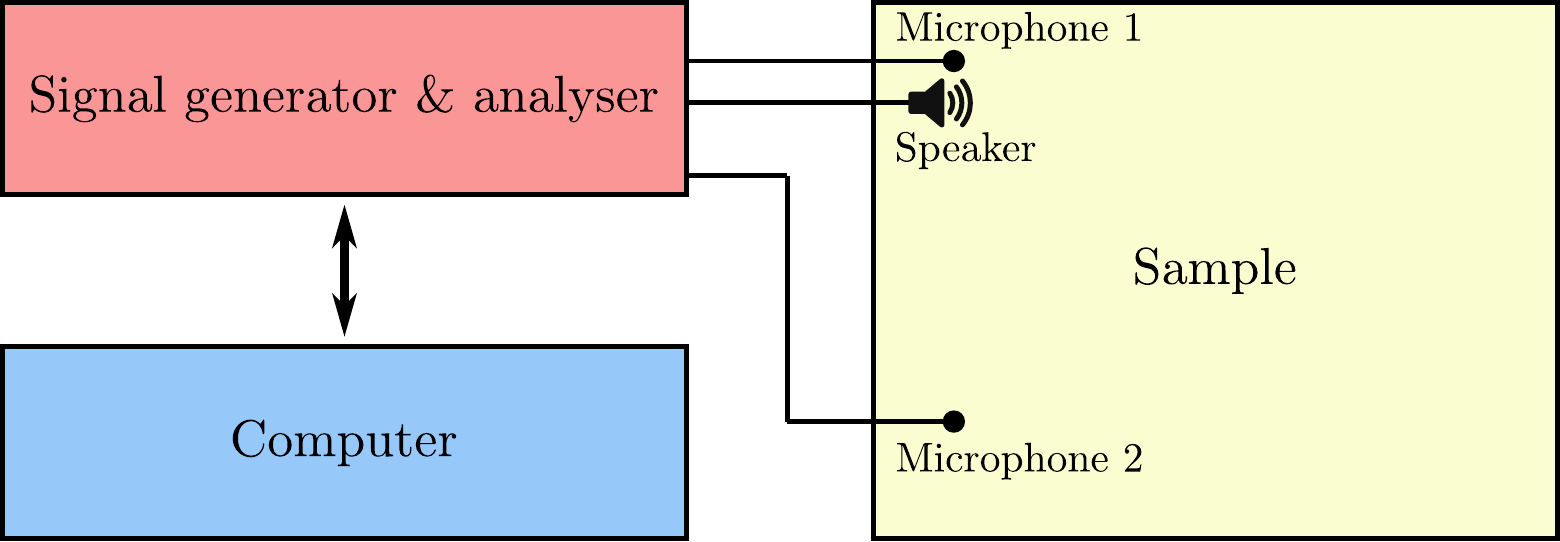}
  \caption{\textbf{Schematic of the experimental setup.} In the experiments, the properties of the excitation signal (i.e., amplitude, frequency range, etc.) and the processing process (i.e., average time, data collecting frequencies, etc.) are set in the computer and passed to the signal generator and analyser. Then, a speaker connected to the signal generator and analyser launches the sound signal into the sample accordingly. Next, the acoustic field distribution in the sample is measured by the microphones. Here, microphone 1 is placed next to the speaker, working as the reference probe. Microphone 2 is the scanning probe and measures the field distribution. Finally, the measured signal is processed by the signal generator and analyser and the processed data are sent to the computer for plotting and further analysis.}
  \label{fige1}
\end{figure*} 

\begin{figure*}
  \centering
  \includegraphics[width=0.8\textwidth]{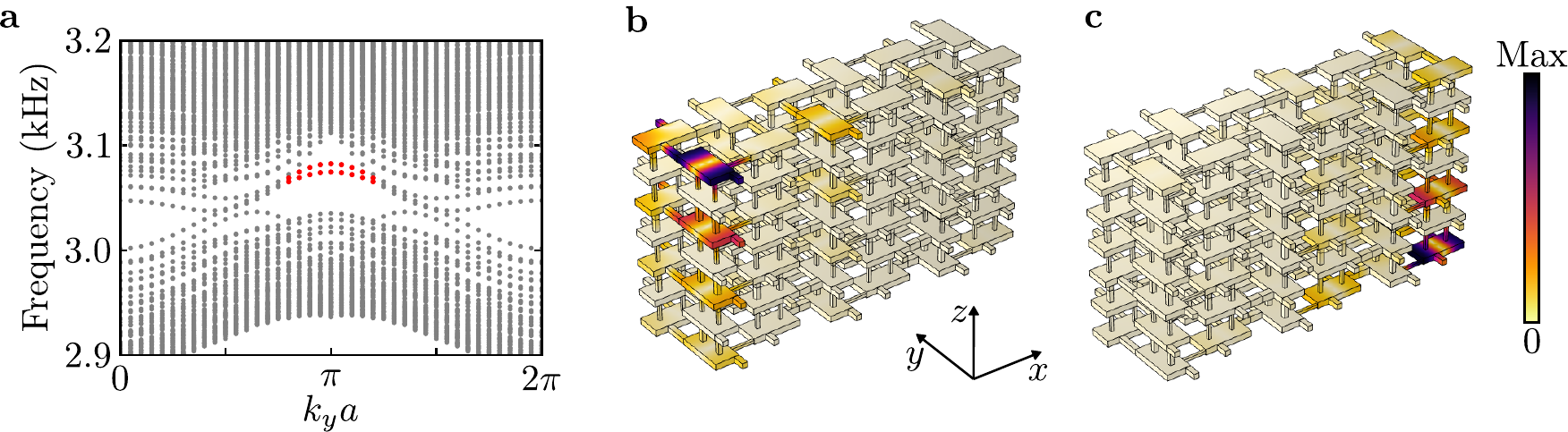}
  \caption{\textbf{Diagonal hinge states by reversing the $z$-dimensional dimerization.} \textbf{a}, Simulated dispersion for a structure similar to the one in Fig.~\ref{fig4} but with swapped $d_{z1}$ and $d_{z2}$. The grey dots represent the bulk and SSs, and the red dots indicate the hinge bands. \textbf{b}-\textbf{c}, Eigen profiles for the two HSs at $k_y=\pi/a$, showing the diagonal distribution of the HSs. The color indicates the amplitude of the acoustic pressure. The corresponding eigenfrequencies are 3072.3 Hz (\textbf{b}) and 3080.8 Hz (\textbf{c}). In \textbf{a}, for a clear visualization of the dispersion, we use 16 and 17 cavities along the $x$ and $z$ directions, respectively. In \textbf{b}-\textbf{c}, we use 6 and 7 cavities along the $x$ and $z$ directions, respectively. Periodic boundary condition is imposed to the $y$ direction for both simulations.}
  \label{fige2}
\end{figure*} 

\begin{figure*}
  \centering
  \includegraphics[width=\textwidth]{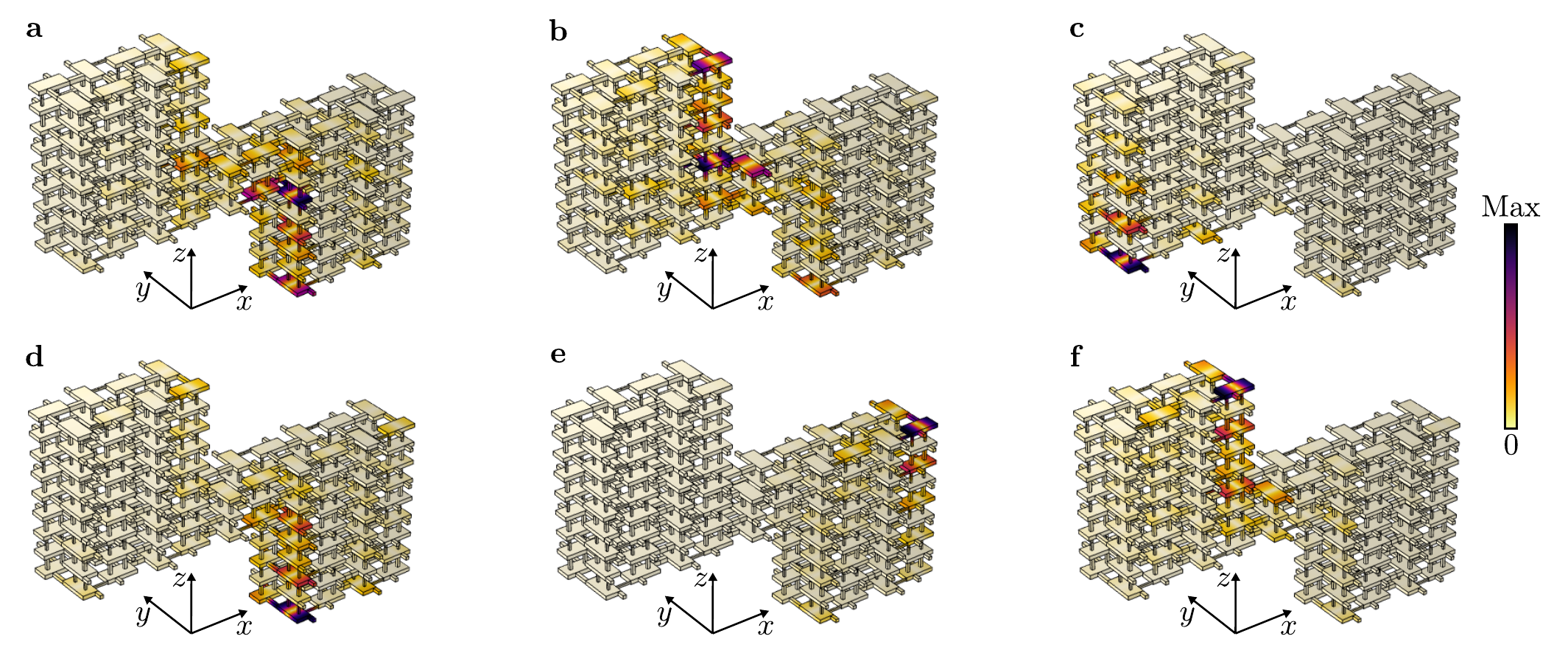}
  \caption{\textbf{Eigen profiles of the six states highlighted in red in Fig.~\ref{fig5}c.} The color indicates the amplitude of the acoustic pressure. The corresponding eigenfrequencies are 3058.3 Hz (\textbf{a}), 3067.4 Hz (\textbf{b}), 3072.6 Hz (\textbf{c}), 3076.4 Hz (\textbf{d}), 3080.6 Hz (\textbf{e}) and 3084.8 Hz (\textbf{f}). These states are localized around the six hinges denoted by the red circles in Fig.~\ref{fig5}b. In particular, the HSs at the two obtuse-angled hinges feature a ``off-center" field pattern (see \textbf{a} and \textbf{b}), consistent with the measurements shown in Extended Data Fig.~\ref{fige5}c, g.}
  \label{fige3}
\end{figure*} 

\begin{figure*}
  \centering
  \includegraphics[width=0.5\textwidth]{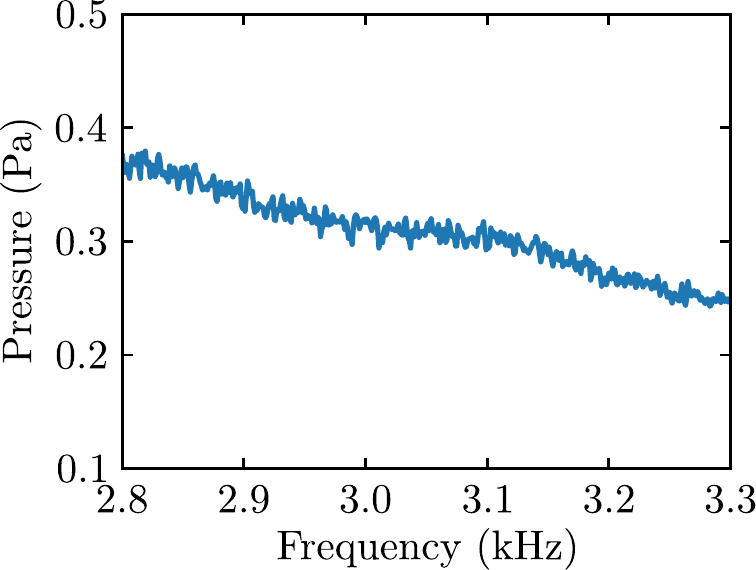}
  \caption{\textbf{Measured acoustic pressure spectrum of the source.}}
  \label{fige4}
\end{figure*} 

\begin{figure*}
  \centering
  \includegraphics[width=\textwidth]{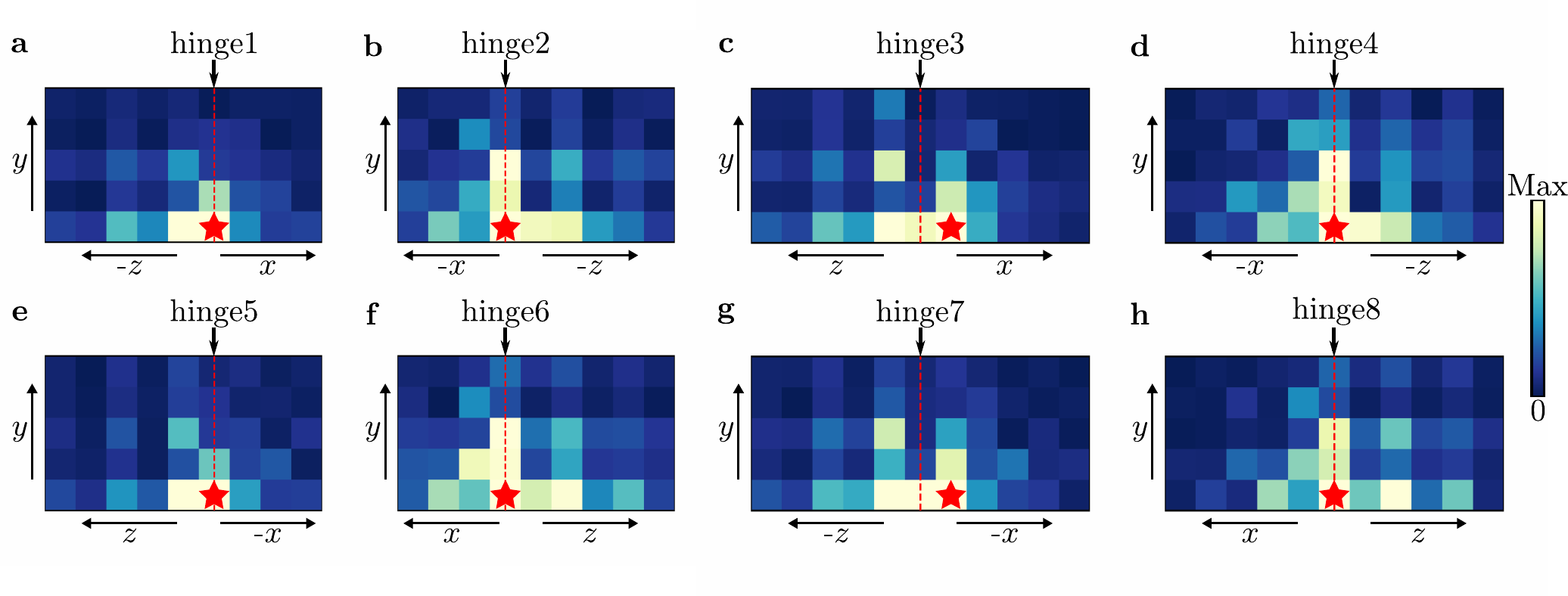}
  \caption{\textbf{Measured acoustic field distributions for the sample shown in Fig.~\ref{fig5}a.} \textbf{a}--\textbf{h}, Experimentally measured acoustic intensity distributions on two surfaces adjacent to hinge 1 (\textbf{a}), hinge 2 (\textbf{b}), hinge 3 (\textbf{c}), hinge 4 (\textbf{d}), hinge 5 (\textbf{e}), hinge 6 (\textbf{f}), hinge 7 (\textbf{g}), hinge 8 (\textbf{h}). The red star indicates the position of the speaker and the red dashed line highlights the position of the hinge. The operating frequencies of the speaker are chosen as: 3075 Hz (hinge 1 and hinge 5), 3076 Hz (hinge 2), 3062 Hz (hinge 3) and 3081 Hz (hinge 4), 3078 Hz (hinge 6), 3051 Hz (hinge 7)  and 3067 Hz (hinge 8),  which are around the eigenfrequencies of the HSs.}
  \label{fige5}
\end{figure*} 


\begin{thebibliography}{40}%
\makeatletter
\providecommand \@ifxundefined [1]{%
 \@ifx{#1\undefined}
}%
\providecommand \@ifnum [1]{%
 \ifnum #1\expandafter \@firstoftwo
 \else \expandafter \@secondoftwo
 \fi
}%
\providecommand \@ifx [1]{%
 \ifx #1\expandafter \@firstoftwo
 \else \expandafter \@secondoftwo
 \fi
}%
\providecommand \natexlab [1]{#1}%
\providecommand \enquote  [1]{``#1''}%
\providecommand \bibnamefont  [1]{#1}%
\providecommand \bibfnamefont [1]{#1}%
\providecommand \citenamefont [1]{#1}%
\providecommand \href@noop [0]{\@secondoftwo}%
\providecommand \href [0]{\begingroup \@sanitize@url \@href}%
\providecommand \@href[1]{\@@startlink{#1}\@@href}%
\providecommand \@@href[1]{\endgroup#1\@@endlink}%
\providecommand \@sanitize@url [0]{\catcode `\\12\catcode `\$12\catcode
  `\&12\catcode `\#12\catcode `\^12\catcode `\_12\catcode `\%12\relax}%
\providecommand \@@startlink[1]{}%
\providecommand \@@endlink[0]{}%
\providecommand \url  [0]{\begingroup\@sanitize@url \@url }%
\providecommand \@url [1]{\endgroup\@href {#1}{\urlprefix }}%
\providecommand \urlprefix  [0]{URL }%
\providecommand \Eprint [0]{\href }%
\providecommand \doibase [0]{https://doi.org/}%
\providecommand \selectlanguage [0]{\@gobble}%
\providecommand \bibinfo  [0]{\@secondoftwo}%
\providecommand \bibfield  [0]{\@secondoftwo}%
\providecommand \translation [1]{[#1]}%
\providecommand \BibitemOpen [0]{}%
\providecommand \bibitemStop [0]{}%
\providecommand \bibitemNoStop [0]{.\EOS\space}%
\providecommand \EOS [0]{\spacefactor3000\relax}%
\providecommand \BibitemShut  [1]{\csname bibitem#1\endcsname}%
\let\auto@bib@innerbib\@empty
%</preamble>
\bibitem [{\citenamefont {Thouless}\ \emph {et~al.}(1982)\citenamefont
  {Thouless}, \citenamefont {Kohmoto}, \citenamefont {Nightingale},\ and\
  \citenamefont {den Nijs}}]{thouless1982}%
  \BibitemOpen
  \bibfield  {author} {\bibinfo {author} {\bibfnamefont {D.~J.}\ \bibnamefont
  {Thouless}}, \bibinfo {author} {\bibfnamefont {M.}~\bibnamefont {Kohmoto}},
  \bibinfo {author} {\bibfnamefont {M.~P.}\ \bibnamefont {Nightingale}},\ and\
  \bibinfo {author} {\bibfnamefont {M.}~\bibnamefont {den Nijs}},\ }\bibfield
  {title} {\bibinfo {title} {Quantized {H}all conductance in a two-dimensional
  periodic potential},\ }\href {https://doi.org/10.1103/PhysRevLett.49.405}
  {\bibfield  {journal} {\bibinfo  {journal} {Phys. Rev. Lett.}\ }\textbf
  {\bibinfo {volume} {49}},\ \bibinfo {pages} {405} (\bibinfo {year}
  {1982})}\BibitemShut {NoStop}%
\bibitem [{\citenamefont {Hasan}\ and\ \citenamefont {Kane}(2010)}]{hasan2010}%
  \BibitemOpen
  \bibfield  {author} {\bibinfo {author} {\bibfnamefont {M.~Z.}\ \bibnamefont
  {Hasan}}\ and\ \bibinfo {author} {\bibfnamefont {C.~L.}\ \bibnamefont
  {Kane}},\ }\bibfield  {title} {\bibinfo {title} {Colloquium: topological
  insulators},\ }\href {https://doi.org/10.1103/RevModPhys.82.3045} {\bibfield
  {journal} {\bibinfo  {journal} {Rev. Mod. Phys.}\ }\textbf {\bibinfo {volume}
  {82}},\ \bibinfo {pages} {3045} (\bibinfo {year} {2010})}\BibitemShut
  {NoStop}%
\bibitem [{\citenamefont {Qi}\ and\ \citenamefont {Zhang}(2011)}]{qi2011}%
  \BibitemOpen
  \bibfield  {author} {\bibinfo {author} {\bibfnamefont {X.-L.}\ \bibnamefont
  {Qi}}\ and\ \bibinfo {author} {\bibfnamefont {S.-C.}\ \bibnamefont {Zhang}},\
  }\bibfield  {title} {\bibinfo {title} {Topological insulators and
  superconductors},\ }\href {https://doi.org/10.1103/RevModPhys.83.1057}
  {\bibfield  {journal} {\bibinfo  {journal} {Rev. Mod. Phys.}\ }\textbf
  {\bibinfo {volume} {83}},\ \bibinfo {pages} {1057} (\bibinfo {year}
  {2011})}\BibitemShut {NoStop}%
\bibitem [{\citenamefont {Haldane}(1988)}]{haldane1988}%
  \BibitemOpen
  \bibfield  {author} {\bibinfo {author} {\bibfnamefont {F.~D.~M.}\
  \bibnamefont {Haldane}},\ }\bibfield  {title} {\bibinfo {title} {Model for a
  quantum {H}all effect without {L}andau levels: {C}ondensed-matter realization
  of the" parity anomaly"},\ }\href
  {https://doi.org/10.1103/PhysRevLett.61.2015} {\bibfield  {journal} {\bibinfo
   {journal} {Phys. Rev. Lett.}\ }\textbf {\bibinfo {volume} {61}},\ \bibinfo
  {pages} {2015} (\bibinfo {year} {1988})}\BibitemShut {NoStop}%
\bibitem [{\citenamefont {Qi}\ \emph {et~al.}(2008)\citenamefont {Qi},
  \citenamefont {Hughes},\ and\ \citenamefont {Zhang}}]{qi2008}%
  \BibitemOpen
  \bibfield  {author} {\bibinfo {author} {\bibfnamefont {X.-L.}\ \bibnamefont
  {Qi}}, \bibinfo {author} {\bibfnamefont {T.~L.}\ \bibnamefont {Hughes}},\
  and\ \bibinfo {author} {\bibfnamefont {S.-C.}\ \bibnamefont {Zhang}},\
  }\bibfield  {title} {\bibinfo {title} {Topological field theory of
  time-reversal invariant insulators},\ }\href
  {https://doi.org/10.1103/PhysRevB.78.195424} {\bibfield  {journal} {\bibinfo
  {journal} {Phys. Rev. B}\ }\textbf {\bibinfo {volume} {78}},\ \bibinfo
  {pages} {195424} (\bibinfo {year} {2008})}\BibitemShut {NoStop}%
\bibitem [{\citenamefont {Wan}\ \emph {et~al.}(2011)\citenamefont {Wan},
  \citenamefont {Turner}, \citenamefont {Vishwanath},\ and\ \citenamefont
  {Savrasov}}]{wan2011}%
  \BibitemOpen
  \bibfield  {author} {\bibinfo {author} {\bibfnamefont {X.}~\bibnamefont
  {Wan}}, \bibinfo {author} {\bibfnamefont {A.~M.}\ \bibnamefont {Turner}},
  \bibinfo {author} {\bibfnamefont {A.}~\bibnamefont {Vishwanath}},\ and\
  \bibinfo {author} {\bibfnamefont {S.~Y.}\ \bibnamefont {Savrasov}},\
  }\bibfield  {title} {\bibinfo {title} {Topological semimetal and {F}ermi-arc
  surface states in the electronic structure of pyrochlore iridates},\ }\href
  {https://doi.org/10.1103/PhysRevB.83.205101} {\bibfield  {journal} {\bibinfo
  {journal} {Phys. Rev. B}\ }\textbf {\bibinfo {volume} {83}},\ \bibinfo
  {pages} {205101} (\bibinfo {year} {2011})}\BibitemShut {NoStop}%
\bibitem [{\citenamefont {Armitage}\ \emph {et~al.}(2018)\citenamefont
  {Armitage}, \citenamefont {Mele},\ and\ \citenamefont
  {Vishwanath}}]{armitage2018}%
  \BibitemOpen
  \bibfield  {author} {\bibinfo {author} {\bibfnamefont {N.}~\bibnamefont
  {Armitage}}, \bibinfo {author} {\bibfnamefont {E.}~\bibnamefont {Mele}},\
  and\ \bibinfo {author} {\bibfnamefont {A.}~\bibnamefont {Vishwanath}},\
  }\bibfield  {title} {\bibinfo {title} {Weyl and {D}irac semimetals in
  three-dimensional solids},\ }\href
  {https://doi.org/10.1103/RevModPhys.90.015001} {\bibfield  {journal}
  {\bibinfo  {journal} {Rev. Mod. Phys.}\ }\textbf {\bibinfo {volume} {90}},\
  \bibinfo {pages} {015001} (\bibinfo {year} {2018})}\BibitemShut {NoStop}%
\bibitem [{\citenamefont {Chiu}\ \emph {et~al.}(2016)\citenamefont {Chiu},
  \citenamefont {Teo}, \citenamefont {Schnyder},\ and\ \citenamefont
  {Ryu}}]{chiu2016}%
  \BibitemOpen
  \bibfield  {author} {\bibinfo {author} {\bibfnamefont {C.-K.}\ \bibnamefont
  {Chiu}}, \bibinfo {author} {\bibfnamefont {J.~C.}\ \bibnamefont {Teo}},
  \bibinfo {author} {\bibfnamefont {A.~P.}\ \bibnamefont {Schnyder}},\ and\
  \bibinfo {author} {\bibfnamefont {S.}~\bibnamefont {Ryu}},\ }\bibfield
  {title} {\bibinfo {title} {Classification of topological quantum matter with
  symmetries},\ }\href {https://doi.org/10.1103/RevModPhys.88.035005}
  {\bibfield  {journal} {\bibinfo  {journal} {Rev. Mod. Phys.}\ }\textbf
  {\bibinfo {volume} {88}},\ \bibinfo {pages} {035005} (\bibinfo {year}
  {2016})}\BibitemShut {NoStop}%
\bibitem [{\citenamefont {Bender}\ and\ \citenamefont
  {Boettcher}(1998)}]{bender1998}%
  \BibitemOpen
  \bibfield  {author} {\bibinfo {author} {\bibfnamefont {C.~M.}\ \bibnamefont
  {Bender}}\ and\ \bibinfo {author} {\bibfnamefont {S.}~\bibnamefont
  {Boettcher}},\ }\bibfield  {title} {\bibinfo {title} {Real spectra in
  non-{H}ermitian {H}amiltonians having {PT} symmetry},\ }\href
  {https://doi.org/10.1103/PhysRevLett.80.5243} {\bibfield  {journal} {\bibinfo
   {journal} {Phys. Rev. Lett.}\ }\textbf {\bibinfo {volume} {80}},\ \bibinfo
  {pages} {5243} (\bibinfo {year} {1998})}\BibitemShut {NoStop}%
\bibitem [{\citenamefont {Zhao}\ \emph {et~al.}(2016)\citenamefont {Zhao},
  \citenamefont {Schnyder},\ and\ \citenamefont {Wang}}]{Zhao2016}%
  \BibitemOpen
  \bibfield  {author} {\bibinfo {author} {\bibfnamefont {Y.~X.}\ \bibnamefont
  {Zhao}}, \bibinfo {author} {\bibfnamefont {A.~P.}\ \bibnamefont {Schnyder}},\
  and\ \bibinfo {author} {\bibfnamefont {Z.~D.}\ \bibnamefont {Wang}},\
  }\bibfield  {title} {\bibinfo {title} {Unified theory of ${PT}$ and ${CP}$
  invariant topological metals and nodal superconductors},\ }\href
  {https://doi.org/10.1103/PhysRevLett.116.156402} {\bibfield  {journal}
  {\bibinfo  {journal} {Phys. Rev. Lett.}\ }\textbf {\bibinfo {volume} {116}},\
  \bibinfo {pages} {156402} (\bibinfo {year} {2016})}\BibitemShut {NoStop}%
\bibitem [{\citenamefont {Zhao}\ and\ \citenamefont {Lu}(2017)}]{Zhao2017}%
  \BibitemOpen
  \bibfield  {author} {\bibinfo {author} {\bibfnamefont {Y.~X.}\ \bibnamefont
  {Zhao}}\ and\ \bibinfo {author} {\bibfnamefont {Y.}~\bibnamefont {Lu}},\
  }\bibfield  {title} {\bibinfo {title} {${PT}$-symmetric real {D}irac fermions
  and semimetals},\ }\href {https://doi.org/10.1103/PhysRevLett.118.056401}
  {\bibfield  {journal} {\bibinfo  {journal} {Phys. Rev. Lett.}\ }\textbf
  {\bibinfo {volume} {118}},\ \bibinfo {pages} {056401} (\bibinfo {year}
  {2017})}\BibitemShut {NoStop}%
\bibitem [{\citenamefont {Nakahara}(2018)}]{nakahara2018geometry}%
  \BibitemOpen
  \bibfield  {author} {\bibinfo {author} {\bibfnamefont {M.}~\bibnamefont
  {Nakahara}},\ }\href@noop {} {\emph {\bibinfo {title} {Geometry, topology and
  physics}}}\ (\bibinfo  {publisher} {CRC press},\ \bibinfo {year}
  {2018})\BibitemShut {NoStop}%
\bibitem [{\citenamefont {Ahn}\ \emph {et~al.}(2018)\citenamefont {Ahn},
  \citenamefont {Kim}, \citenamefont {Kim},\ and\ \citenamefont
  {Yang}}]{Yang_2018prl}%
  \BibitemOpen
  \bibfield  {author} {\bibinfo {author} {\bibfnamefont {J.}~\bibnamefont
  {Ahn}}, \bibinfo {author} {\bibfnamefont {D.}~\bibnamefont {Kim}}, \bibinfo
  {author} {\bibfnamefont {Y.}~\bibnamefont {Kim}},\ and\ \bibinfo {author}
  {\bibfnamefont {B.-J.}\ \bibnamefont {Yang}},\ }\bibfield  {title} {\bibinfo
  {title} {Band topology and linking structure of nodal line semimetals with
  ${Z}_{2}$ monopole charges},\ }\href
  {https://doi.org/10.1103/PhysRevLett.121.106403} {\bibfield  {journal}
  {\bibinfo  {journal} {Phys. Rev. Lett.}\ }\textbf {\bibinfo {volume} {121}},\
  \bibinfo {pages} {106403} (\bibinfo {year} {2018})}\BibitemShut {NoStop}%
\bibitem [{\citenamefont {Ahn}\ \emph {et~al.}(2019)\citenamefont {Ahn},
  \citenamefont {Park},\ and\ \citenamefont {Yang}}]{Yang_2019prx}%
  \BibitemOpen
  \bibfield  {author} {\bibinfo {author} {\bibfnamefont {J.}~\bibnamefont
  {Ahn}}, \bibinfo {author} {\bibfnamefont {S.}~\bibnamefont {Park}},\ and\
  \bibinfo {author} {\bibfnamefont {B.-J.}\ \bibnamefont {Yang}},\ }\bibfield
  {title} {\bibinfo {title} {Failure of {N}ielsen-{N}inomiya theorem and
  fragile topology in two-dimensional systems with space-time inversion
  symmetry: Application to twisted bilayer graphene at magic angle},\ }\href
  {https://doi.org/10.1103/PhysRevX.9.021013} {\bibfield  {journal} {\bibinfo
  {journal} {Phys. Rev. X}\ }\textbf {\bibinfo {volume} {9}},\ \bibinfo {pages}
  {021013} (\bibinfo {year} {2019})}\BibitemShut {NoStop}%
\bibitem [{\citenamefont {Bzdu\ifmmode~\check{s}\else \v{s}\fi{}ek}\ and\
  \citenamefont {Sigrist}(2017)}]{Sigrist_2017prb}%
  \BibitemOpen
  \bibfield  {author} {\bibinfo {author} {\bibfnamefont {T.~c.~v.}\
  \bibnamefont {Bzdu\ifmmode~\check{s}\else \v{s}\fi{}ek}}\ and\ \bibinfo
  {author} {\bibfnamefont {M.}~\bibnamefont {Sigrist}},\ }\bibfield  {title}
  {\bibinfo {title} {Robust doubly charged nodal lines and nodal surfaces in
  centrosymmetric systems},\ }\href
  {https://doi.org/10.1103/PhysRevB.96.155105} {\bibfield  {journal} {\bibinfo
  {journal} {Phys. Rev. B}\ }\textbf {\bibinfo {volume} {96}},\ \bibinfo
  {pages} {155105} (\bibinfo {year} {2017})}\BibitemShut {NoStop}%
\bibitem [{\citenamefont {Sheng}\ \emph {et~al.}(2019)\citenamefont {Sheng},
  \citenamefont {Chen}, \citenamefont {Liu}, \citenamefont {Chen},
  \citenamefont {Yu}, \citenamefont {Zhao},\ and\ \citenamefont
  {Yang}}]{Sheng2019}%
  \BibitemOpen
  \bibfield  {author} {\bibinfo {author} {\bibfnamefont {X.-L.}\ \bibnamefont
  {Sheng}}, \bibinfo {author} {\bibfnamefont {C.}~\bibnamefont {Chen}},
  \bibinfo {author} {\bibfnamefont {H.}~\bibnamefont {Liu}}, \bibinfo {author}
  {\bibfnamefont {Z.}~\bibnamefont {Chen}}, \bibinfo {author} {\bibfnamefont
  {Z.-M.}\ \bibnamefont {Yu}}, \bibinfo {author} {\bibfnamefont {Y.~X.}\
  \bibnamefont {Zhao}},\ and\ \bibinfo {author} {\bibfnamefont {S.~A.}\
  \bibnamefont {Yang}},\ }\bibfield  {title} {\bibinfo {title} {Two-dimensional
  second-order topological insulator in graphdiyne},\ }\href
  {https://doi.org/10.1103/PhysRevLett.123.256402} {\bibfield  {journal}
  {\bibinfo  {journal} {Phys. Rev. Lett.}\ }\textbf {\bibinfo {volume} {123}},\
  \bibinfo {pages} {256402} (\bibinfo {year} {2019})}\BibitemShut {NoStop}%
\bibitem [{\citenamefont {Wang}\ \emph {et~al.}(2020)\citenamefont {Wang},
  \citenamefont {Dai}, \citenamefont {Shao}, \citenamefont {Yang},\ and\
  \citenamefont {Zhao}}]{wang2020}%
  \BibitemOpen
  \bibfield  {author} {\bibinfo {author} {\bibfnamefont {K.}~\bibnamefont
  {Wang}}, \bibinfo {author} {\bibfnamefont {J.-X.}\ \bibnamefont {Dai}},
  \bibinfo {author} {\bibfnamefont {L.}~\bibnamefont {Shao}}, \bibinfo {author}
  {\bibfnamefont {S.~A.}\ \bibnamefont {Yang}},\ and\ \bibinfo {author}
  {\bibfnamefont {Y.}~\bibnamefont {Zhao}},\ }\bibfield  {title} {\bibinfo
  {title} {Boundary criticality of {PT}-invariant topology and second-order
  nodal-line semimetals},\ }\href
  {https://doi.org/10.1103/PhysRevLett.125.126403} {\bibfield  {journal}
  {\bibinfo  {journal} {Phys. Rev. Lett.}\ }\textbf {\bibinfo {volume} {125}},\
  \bibinfo {pages} {126403} (\bibinfo {year} {2020})}\BibitemShut {NoStop}%
\bibitem [{\citenamefont {Chen}\ \emph {et~al.}(2022)\citenamefont {Chen},
  \citenamefont {Zeng}, \citenamefont {Chen}, \citenamefont {Zhao},
  \citenamefont {Sheng},\ and\ \citenamefont {Yang}}]{Sheng2022}%
  \BibitemOpen
  \bibfield  {author} {\bibinfo {author} {\bibfnamefont {C.}~\bibnamefont
  {Chen}}, \bibinfo {author} {\bibfnamefont {X.-T.}\ \bibnamefont {Zeng}},
  \bibinfo {author} {\bibfnamefont {Z.}~\bibnamefont {Chen}}, \bibinfo {author}
  {\bibfnamefont {Y.~X.}\ \bibnamefont {Zhao}}, \bibinfo {author}
  {\bibfnamefont {X.-L.}\ \bibnamefont {Sheng}},\ and\ \bibinfo {author}
  {\bibfnamefont {S.~A.}\ \bibnamefont {Yang}},\ }\bibfield  {title} {\bibinfo
  {title} {Second-order real nodal-line semimetal in three-dimensional
  graphdiyne},\ }\href {https://doi.org/10.1103/PhysRevLett.128.026405}
  {\bibfield  {journal} {\bibinfo  {journal} {Phys. Rev. Lett.}\ }\textbf
  {\bibinfo {volume} {128}},\ \bibinfo {pages} {026405} (\bibinfo {year}
  {2022})}\BibitemShut {NoStop}%
\bibitem [{\citenamefont {Nielsen}\ and\ \citenamefont
  {Ninomiya}(1981)}]{nielsen1981absence}%
  \BibitemOpen
  \bibfield  {author} {\bibinfo {author} {\bibfnamefont {H.~B.}\ \bibnamefont
  {Nielsen}}\ and\ \bibinfo {author} {\bibfnamefont {M.}~\bibnamefont
  {Ninomiya}},\ }\bibfield  {title} {\bibinfo {title} {Absence of neutrinos on
  a lattice:(i). proof by homotopy theory},\ }\href
  {https://doi.org/10.1016/0550-3213(81)90361-8} {\bibfield  {journal}
  {\bibinfo  {journal} {Nuclear Physics B}\ }\textbf {\bibinfo {volume}
  {185}},\ \bibinfo {pages} {20} (\bibinfo {year} {1981})}\BibitemShut
  {NoStop}%
\bibitem [{\citenamefont {Weng}\ \emph {et~al.}(2015)\citenamefont {Weng},
  \citenamefont {Liang}, \citenamefont {Xu}, \citenamefont {Yu}, \citenamefont
  {Fang}, \citenamefont {Dai},\ and\ \citenamefont {Kawazoe}}]{weng2015}%
  \BibitemOpen
  \bibfield  {author} {\bibinfo {author} {\bibfnamefont {H.}~\bibnamefont
  {Weng}}, \bibinfo {author} {\bibfnamefont {Y.}~\bibnamefont {Liang}},
  \bibinfo {author} {\bibfnamefont {Q.}~\bibnamefont {Xu}}, \bibinfo {author}
  {\bibfnamefont {R.}~\bibnamefont {Yu}}, \bibinfo {author} {\bibfnamefont
  {Z.}~\bibnamefont {Fang}}, \bibinfo {author} {\bibfnamefont {X.}~\bibnamefont
  {Dai}},\ and\ \bibinfo {author} {\bibfnamefont {Y.}~\bibnamefont {Kawazoe}},\
  }\bibfield  {title} {\bibinfo {title} {Topological node-line semimetal in
  three-dimensional graphene networks},\ }\href
  {https://doi.org/10.1103/PhysRevB.92.045108} {\bibfield  {journal} {\bibinfo
  {journal} {Phys. Rev. B}\ }\textbf {\bibinfo {volume} {92}},\ \bibinfo
  {pages} {045108} (\bibinfo {year} {2015})}\BibitemShut {NoStop}%
\bibitem [{\citenamefont {Luo}\ \emph {et~al.}(2021)\citenamefont {Luo},
  \citenamefont {Wang}, \citenamefont {Lin}, \citenamefont {Jiang},
  \citenamefont {Wu}, \citenamefont {Li},\ and\ \citenamefont
  {Jiang}}]{luo2021}%
  \BibitemOpen
  \bibfield  {author} {\bibinfo {author} {\bibfnamefont {L.}~\bibnamefont
  {Luo}}, \bibinfo {author} {\bibfnamefont {H.-X.}\ \bibnamefont {Wang}},
  \bibinfo {author} {\bibfnamefont {Z.-K.}\ \bibnamefont {Lin}}, \bibinfo
  {author} {\bibfnamefont {B.}~\bibnamefont {Jiang}}, \bibinfo {author}
  {\bibfnamefont {Y.}~\bibnamefont {Wu}}, \bibinfo {author} {\bibfnamefont
  {F.}~\bibnamefont {Li}},\ and\ \bibinfo {author} {\bibfnamefont {J.-H.}\
  \bibnamefont {Jiang}},\ }\bibfield  {title} {\bibinfo {title} {Observation of
  a phononic higher-order {W}eyl semimetal},\ }\href
  {https://doi.org/10.1038/s41563-021-00985-6} {\bibfield  {journal} {\bibinfo
  {journal} {Nat. Mater.}\ }\textbf {\bibinfo {volume} {20}},\ \bibinfo {pages}
  {794} (\bibinfo {year} {2021})}\BibitemShut {NoStop}%
\bibitem [{\citenamefont {Wei}\ \emph {et~al.}(2021)\citenamefont {Wei},
  \citenamefont {Zhang}, \citenamefont {Deng}, \citenamefont {Lu},
  \citenamefont {Huang}, \citenamefont {Yan}, \citenamefont {Chen},
  \citenamefont {Liu},\ and\ \citenamefont {Jia}}]{wei2021}%
  \BibitemOpen
  \bibfield  {author} {\bibinfo {author} {\bibfnamefont {Q.}~\bibnamefont
  {Wei}}, \bibinfo {author} {\bibfnamefont {X.}~\bibnamefont {Zhang}}, \bibinfo
  {author} {\bibfnamefont {W.}~\bibnamefont {Deng}}, \bibinfo {author}
  {\bibfnamefont {J.}~\bibnamefont {Lu}}, \bibinfo {author} {\bibfnamefont
  {X.}~\bibnamefont {Huang}}, \bibinfo {author} {\bibfnamefont
  {M.}~\bibnamefont {Yan}}, \bibinfo {author} {\bibfnamefont {G.}~\bibnamefont
  {Chen}}, \bibinfo {author} {\bibfnamefont {Z.}~\bibnamefont {Liu}},\ and\
  \bibinfo {author} {\bibfnamefont {S.}~\bibnamefont {Jia}},\ }\bibfield
  {title} {\bibinfo {title} {Higher-order topological semimetal in acoustic
  crystals},\ }\href {https://doi.org/10.1038/s41563-021-00933-4} {\bibfield
  {journal} {\bibinfo  {journal} {Nat. Mater.}\ }\textbf {\bibinfo {volume}
  {20}},\ \bibinfo {pages} {812} (\bibinfo {year} {2021})}\BibitemShut
  {NoStop}%
\bibitem [{\citenamefont {Qiu}\ \emph {et~al.}(2021)\citenamefont {Qiu},
  \citenamefont {Xiao}, \citenamefont {Zhang},\ and\ \citenamefont
  {Qiu}}]{qiu2021}%
  \BibitemOpen
  \bibfield  {author} {\bibinfo {author} {\bibfnamefont {H.}~\bibnamefont
  {Qiu}}, \bibinfo {author} {\bibfnamefont {M.}~\bibnamefont {Xiao}}, \bibinfo
  {author} {\bibfnamefont {F.}~\bibnamefont {Zhang}},\ and\ \bibinfo {author}
  {\bibfnamefont {C.}~\bibnamefont {Qiu}},\ }\bibfield  {title} {\bibinfo
  {title} {Higher-order {D}irac sonic crystals},\ }\href
  {https://doi.org/10.1103/PhysRevLett.127.146601} {\bibfield  {journal}
  {\bibinfo  {journal} {Phys. Rev. Lett.}\ }\textbf {\bibinfo {volume} {127}},\
  \bibinfo {pages} {146601} (\bibinfo {year} {2021})}\BibitemShut {NoStop}%
\bibitem [{\citenamefont {Wen}(2002)}]{Wen-PSG}%
  \BibitemOpen
  \bibfield  {author} {\bibinfo {author} {\bibfnamefont {X.-G.}\ \bibnamefont
  {Wen}},\ }\bibfield  {title} {\bibinfo {title} {Quantum orders and symmetric
  spin liquids},\ }\href {https://doi.org/10.1103/PhysRevB.65.165113}
  {\bibfield  {journal} {\bibinfo  {journal} {Phys. Rev. B}\ }\textbf {\bibinfo
  {volume} {65}},\ \bibinfo {pages} {165113} (\bibinfo {year}
  {2002})}\BibitemShut {NoStop}%
\bibitem [{\citenamefont {Zhao}\ \emph {et~al.}(2020)\citenamefont {Zhao},
  \citenamefont {Huang},\ and\ \citenamefont {Yang}}]{Zhao_2020prb}%
  \BibitemOpen
  \bibfield  {author} {\bibinfo {author} {\bibfnamefont {Y.~X.}\ \bibnamefont
  {Zhao}}, \bibinfo {author} {\bibfnamefont {Y.-X.}\ \bibnamefont {Huang}},\
  and\ \bibinfo {author} {\bibfnamefont {S.~A.}\ \bibnamefont {Yang}},\
  }\bibfield  {title} {\bibinfo {title} {$\mathbb{Z}_2$-projective
  translational symmetry protected topological phases},\ }\href
  {https://doi.org/10.1103/PhysRevB.102.161117} {\bibfield  {journal} {\bibinfo
   {journal} {Phys. Rev. B}\ }\textbf {\bibinfo {volume} {102}},\ \bibinfo
  {pages} {161117} (\bibinfo {year} {2020})}\BibitemShut {NoStop}%
\bibitem [{\citenamefont {Shao}\ \emph {et~al.}(2021)\citenamefont {Shao},
  \citenamefont {Liu}, \citenamefont {Xiao}, \citenamefont {Yang},\ and\
  \citenamefont {Zhao}}]{Shao_2021prl}%
  \BibitemOpen
  \bibfield  {author} {\bibinfo {author} {\bibfnamefont {L.~B.}\ \bibnamefont
  {Shao}}, \bibinfo {author} {\bibfnamefont {Q.}~\bibnamefont {Liu}}, \bibinfo
  {author} {\bibfnamefont {R.}~\bibnamefont {Xiao}}, \bibinfo {author}
  {\bibfnamefont {S.~A.}\ \bibnamefont {Yang}},\ and\ \bibinfo {author}
  {\bibfnamefont {Y.~X.}\ \bibnamefont {Zhao}},\ }\bibfield  {title} {\bibinfo
  {title} {Gauge-field extended $k\ifmmode\cdot\else\textperiodcentered\fi{}p$
  method and novel topological phases},\ }\href
  {https://doi.org/10.1103/PhysRevLett.127.076401} {\bibfield  {journal}
  {\bibinfo  {journal} {Phys. Rev. Lett.}\ }\textbf {\bibinfo {volume} {127}},\
  \bibinfo {pages} {076401} (\bibinfo {year} {2021})}\BibitemShut {NoStop}%
\bibitem [{\citenamefont {Xue}\ \emph {et~al.}(2022{\natexlab{a}})\citenamefont
  {Xue}, \citenamefont {Wang}, \citenamefont {Huang}, \citenamefont {Cheng},
  \citenamefont {Yu}, \citenamefont {Foo}, \citenamefont {Zhao}, \citenamefont
  {Yang},\ and\ \citenamefont {Zhang}}]{xue2022}%
  \BibitemOpen
  \bibfield  {author} {\bibinfo {author} {\bibfnamefont {H.}~\bibnamefont
  {Xue}}, \bibinfo {author} {\bibfnamefont {Z.}~\bibnamefont {Wang}}, \bibinfo
  {author} {\bibfnamefont {Y.-X.}\ \bibnamefont {Huang}}, \bibinfo {author}
  {\bibfnamefont {Z.}~\bibnamefont {Cheng}}, \bibinfo {author} {\bibfnamefont
  {L.}~\bibnamefont {Yu}}, \bibinfo {author} {\bibfnamefont {Y.}~\bibnamefont
  {Foo}}, \bibinfo {author} {\bibfnamefont {Y.}~\bibnamefont {Zhao}}, \bibinfo
  {author} {\bibfnamefont {S.~A.}\ \bibnamefont {Yang}},\ and\ \bibinfo
  {author} {\bibfnamefont {B.}~\bibnamefont {Zhang}},\ }\bibfield  {title}
  {\bibinfo {title} {Projectively enriched symmetry and topology in acoustic
  crystals},\ }\href {https://doi.org/10.1103/PhysRevLett.128.116802}
  {\bibfield  {journal} {\bibinfo  {journal} {Phys. Rev. Lett.}\ }\textbf
  {\bibinfo {volume} {128}},\ \bibinfo {pages} {116802} (\bibinfo {year}
  {2022}{\natexlab{a}})}\BibitemShut {NoStop}%
\bibitem [{\citenamefont {Li}\ \emph {et~al.}(2022)\citenamefont {Li},
  \citenamefont {Du}, \citenamefont {Zhang}, \citenamefont {Li}, \citenamefont
  {Fan}, \citenamefont {Zhang},\ and\ \citenamefont {Qiu}}]{li2022}%
  \BibitemOpen
  \bibfield  {author} {\bibinfo {author} {\bibfnamefont {T.}~\bibnamefont
  {Li}}, \bibinfo {author} {\bibfnamefont {J.}~\bibnamefont {Du}}, \bibinfo
  {author} {\bibfnamefont {Q.}~\bibnamefont {Zhang}}, \bibinfo {author}
  {\bibfnamefont {Y.}~\bibnamefont {Li}}, \bibinfo {author} {\bibfnamefont
  {X.}~\bibnamefont {Fan}}, \bibinfo {author} {\bibfnamefont {F.}~\bibnamefont
  {Zhang}},\ and\ \bibinfo {author} {\bibfnamefont {C.}~\bibnamefont {Qiu}},\
  }\bibfield  {title} {\bibinfo {title} {Acoustic {M}{\"o}bius insulators from
  projective symmetry},\ }\href
  {https://doi.org/10.1103/PhysRevLett.128.116803} {\bibfield  {journal}
  {\bibinfo  {journal} {Phys. Rev. Lett.}\ }\textbf {\bibinfo {volume} {128}},\
  \bibinfo {pages} {116803} (\bibinfo {year} {2022})}\BibitemShut {NoStop}%
\bibitem [{\citenamefont {Ma}\ \emph {et~al.}(2019)\citenamefont {Ma},
  \citenamefont {Xiao},\ and\ \citenamefont {Chan}}]{ma2019}%
  \BibitemOpen
  \bibfield  {author} {\bibinfo {author} {\bibfnamefont {G.}~\bibnamefont
  {Ma}}, \bibinfo {author} {\bibfnamefont {M.}~\bibnamefont {Xiao}},\ and\
  \bibinfo {author} {\bibfnamefont {C.~T.}\ \bibnamefont {Chan}},\ }\bibfield
  {title} {\bibinfo {title} {Topological phases in acoustic and mechanical
  systems},\ }\href {https://doi.org/10.1038/s42254-019-0030-x} {\bibfield
  {journal} {\bibinfo  {journal} {Nat. Rev. Phys.}\ }\textbf {\bibinfo {volume}
  {1}},\ \bibinfo {pages} {281} (\bibinfo {year} {2019})}\BibitemShut {NoStop}%
\bibitem [{\citenamefont {Xue}\ \emph {et~al.}(2022{\natexlab{b}})\citenamefont
  {Xue}, \citenamefont {Yang},\ and\ \citenamefont {Zhang}}]{xue2022b}%
  \BibitemOpen
  \bibfield  {author} {\bibinfo {author} {\bibfnamefont {H.}~\bibnamefont
  {Xue}}, \bibinfo {author} {\bibfnamefont {Y.}~\bibnamefont {Yang}},\ and\
  \bibinfo {author} {\bibfnamefont {B.}~\bibnamefont {Zhang}},\ }\bibfield
  {title} {\bibinfo {title} {Topological acoustics},\ }\href
  {https://doi.org/10.1038/s41578-022-00465-6} {\bibfield  {journal} {\bibinfo
  {journal} {Nat. Rev. Mater.}\ }\textbf {\bibinfo {volume} {7}},\ \bibinfo
  {pages} {974} (\bibinfo {year} {2022}{\natexlab{b}})}\BibitemShut {NoStop}%
\bibitem [{\citenamefont {Xue}\ \emph {et~al.}(2019)\citenamefont {Xue},
  \citenamefont {Yang}, \citenamefont {Gao}, \citenamefont {Chong},\ and\
  \citenamefont {Zhang}}]{xue2019}%
  \BibitemOpen
  \bibfield  {author} {\bibinfo {author} {\bibfnamefont {H.}~\bibnamefont
  {Xue}}, \bibinfo {author} {\bibfnamefont {Y.}~\bibnamefont {Yang}}, \bibinfo
  {author} {\bibfnamefont {F.}~\bibnamefont {Gao}}, \bibinfo {author}
  {\bibfnamefont {Y.}~\bibnamefont {Chong}},\ and\ \bibinfo {author}
  {\bibfnamefont {B.}~\bibnamefont {Zhang}},\ }\bibfield  {title} {\bibinfo
  {title} {Acoustic higher-order topological insulator on a kagome lattice},\
  }\href {https://doi.org/10.1038/s41563-018-0251-x} {\bibfield  {journal}
  {\bibinfo  {journal} {Nat. Mater.}\ }\textbf {\bibinfo {volume} {18}},\
  \bibinfo {pages} {108} (\bibinfo {year} {2019})}\BibitemShut {NoStop}%
\bibitem [{\citenamefont {Ni}\ \emph {et~al.}(2019)\citenamefont {Ni},
  \citenamefont {Weiner}, \citenamefont {Alu},\ and\ \citenamefont
  {Khanikaev}}]{ni2019}%
  \BibitemOpen
  \bibfield  {author} {\bibinfo {author} {\bibfnamefont {X.}~\bibnamefont
  {Ni}}, \bibinfo {author} {\bibfnamefont {M.}~\bibnamefont {Weiner}}, \bibinfo
  {author} {\bibfnamefont {A.}~\bibnamefont {Alu}},\ and\ \bibinfo {author}
  {\bibfnamefont {A.~B.}\ \bibnamefont {Khanikaev}},\ }\bibfield  {title}
  {\bibinfo {title} {Observation of higher-order topological acoustic states
  protected by generalized chiral symmetry},\ }\href
  {https://doi.org/10.1038/s41563-018-0252-9} {\bibfield  {journal} {\bibinfo
  {journal} {Nat. Mater.}\ }\textbf {\bibinfo {volume} {18}},\ \bibinfo {pages}
  {113} (\bibinfo {year} {2019})}\BibitemShut {NoStop}%
\bibitem [{\citenamefont {Li}\ \emph {et~al.}(2018)\citenamefont {Li},
  \citenamefont {Huang}, \citenamefont {Lu}, \citenamefont {Ma},\ and\
  \citenamefont {Liu}}]{li2018}%
  \BibitemOpen
  \bibfield  {author} {\bibinfo {author} {\bibfnamefont {F.}~\bibnamefont
  {Li}}, \bibinfo {author} {\bibfnamefont {X.}~\bibnamefont {Huang}}, \bibinfo
  {author} {\bibfnamefont {J.}~\bibnamefont {Lu}}, \bibinfo {author}
  {\bibfnamefont {J.}~\bibnamefont {Ma}},\ and\ \bibinfo {author}
  {\bibfnamefont {Z.}~\bibnamefont {Liu}},\ }\bibfield  {title} {\bibinfo
  {title} {Weyl points and {F}ermi arcs in a chiral phononic crystal},\ }\href
  {https://doi.org/10.1038/nphys4275} {\bibfield  {journal} {\bibinfo
  {journal} {Nat. Phys.}\ }\textbf {\bibinfo {volume} {14}},\ \bibinfo {pages}
  {30} (\bibinfo {year} {2018})}\BibitemShut {NoStop}%
\bibitem [{\citenamefont {Xue}\ \emph {et~al.}(2020)\citenamefont {Xue},
  \citenamefont {Ge}, \citenamefont {Sun}, \citenamefont {Wang}, \citenamefont
  {Jia}, \citenamefont {Guan}, \citenamefont {Yuan}, \citenamefont {Chong},\
  and\ \citenamefont {Zhang}}]{xue2020}%
  \BibitemOpen
  \bibfield  {author} {\bibinfo {author} {\bibfnamefont {H.}~\bibnamefont
  {Xue}}, \bibinfo {author} {\bibfnamefont {Y.}~\bibnamefont {Ge}}, \bibinfo
  {author} {\bibfnamefont {H.-X.}\ \bibnamefont {Sun}}, \bibinfo {author}
  {\bibfnamefont {Q.}~\bibnamefont {Wang}}, \bibinfo {author} {\bibfnamefont
  {D.}~\bibnamefont {Jia}}, \bibinfo {author} {\bibfnamefont {Y.-J.}\
  \bibnamefont {Guan}}, \bibinfo {author} {\bibfnamefont {S.-Q.}\ \bibnamefont
  {Yuan}}, \bibinfo {author} {\bibfnamefont {Y.}~\bibnamefont {Chong}},\ and\
  \bibinfo {author} {\bibfnamefont {B.}~\bibnamefont {Zhang}},\ }\bibfield
  {title} {\bibinfo {title} {Observation of an acoustic octupole topological
  insulator},\ }\href {https://doi.org/10.1038/s41467-020-16350-1} {\bibfield
  {journal} {\bibinfo  {journal} {Nat. Commun.}\ }\textbf {\bibinfo {volume}
  {11}},\ \bibinfo {pages} {2442} (\bibinfo {year} {2020})}\BibitemShut
  {NoStop}%
\bibitem [{\citenamefont {Ni}\ \emph {et~al.}(2020)\citenamefont {Ni},
  \citenamefont {Li}, \citenamefont {Weiner}, \citenamefont {Al{\`u}},\ and\
  \citenamefont {Khanikaev}}]{ni2020}%
  \BibitemOpen
  \bibfield  {author} {\bibinfo {author} {\bibfnamefont {X.}~\bibnamefont
  {Ni}}, \bibinfo {author} {\bibfnamefont {M.}~\bibnamefont {Li}}, \bibinfo
  {author} {\bibfnamefont {M.}~\bibnamefont {Weiner}}, \bibinfo {author}
  {\bibfnamefont {A.}~\bibnamefont {Al{\`u}}},\ and\ \bibinfo {author}
  {\bibfnamefont {A.~B.}\ \bibnamefont {Khanikaev}},\ }\bibfield  {title}
  {\bibinfo {title} {Demonstration of a quantized acoustic octupole topological
  insulator},\ }\href {https://doi.org/10.1038/s41467-020-15705-y} {\bibfield
  {journal} {\bibinfo  {journal} {Nat. Commun.}\ }\textbf {\bibinfo {volume}
  {11}},\ \bibinfo {pages} {2108} (\bibinfo {year} {2020})}\BibitemShut
  {NoStop}%
\bibitem [{\citenamefont {Qi}\ \emph {et~al.}(2020)\citenamefont {Qi},
  \citenamefont {Qiu}, \citenamefont {Xiao}, \citenamefont {He}, \citenamefont
  {Ke},\ and\ \citenamefont {Liu}}]{qi2020}%
  \BibitemOpen
  \bibfield  {author} {\bibinfo {author} {\bibfnamefont {Y.}~\bibnamefont
  {Qi}}, \bibinfo {author} {\bibfnamefont {C.}~\bibnamefont {Qiu}}, \bibinfo
  {author} {\bibfnamefont {M.}~\bibnamefont {Xiao}}, \bibinfo {author}
  {\bibfnamefont {H.}~\bibnamefont {He}}, \bibinfo {author} {\bibfnamefont
  {M.}~\bibnamefont {Ke}},\ and\ \bibinfo {author} {\bibfnamefont
  {Z.}~\bibnamefont {Liu}},\ }\bibfield  {title} {\bibinfo {title} {Acoustic
  realization of quadrupole topological insulators},\ }\href
  {https://doi.org/10.1103/PhysRevLett.124.206601} {\bibfield  {journal}
  {\bibinfo  {journal} {Phys. Rev. Lett.}\ }\textbf {\bibinfo {volume} {124}},\
  \bibinfo {pages} {206601} (\bibinfo {year} {2020})}\BibitemShut {NoStop}%
\bibitem [{\citenamefont {Serra-Garcia}\ \emph {et~al.}(2018)\citenamefont
  {Serra-Garcia}, \citenamefont {Peri}, \citenamefont {S{\"u}sstrunk},
  \citenamefont {Bilal}, \citenamefont {Larsen}, \citenamefont {Villanueva},\
  and\ \citenamefont {Huber}}]{serra2018}%
  \BibitemOpen
  \bibfield  {author} {\bibinfo {author} {\bibfnamefont {M.}~\bibnamefont
  {Serra-Garcia}}, \bibinfo {author} {\bibfnamefont {V.}~\bibnamefont {Peri}},
  \bibinfo {author} {\bibfnamefont {R.}~\bibnamefont {S{\"u}sstrunk}}, \bibinfo
  {author} {\bibfnamefont {O.~R.}\ \bibnamefont {Bilal}}, \bibinfo {author}
  {\bibfnamefont {T.}~\bibnamefont {Larsen}}, \bibinfo {author} {\bibfnamefont
  {L.~G.}\ \bibnamefont {Villanueva}},\ and\ \bibinfo {author} {\bibfnamefont
  {S.~D.}\ \bibnamefont {Huber}},\ }\bibfield  {title} {\bibinfo {title}
  {Observation of a phononic quadrupole topological insulator},\ }\href
  {https://doi.org/10.1038/nature25156} {\bibfield  {journal} {\bibinfo
  {journal} {Nature}\ }\textbf {\bibinfo {volume} {555}},\ \bibinfo {pages}
  {342} (\bibinfo {year} {2018})}\BibitemShut {NoStop}%
\bibitem [{\citenamefont {Peterson}\ \emph {et~al.}(2018)\citenamefont
  {Peterson}, \citenamefont {Benalcazar}, \citenamefont {Hughes},\ and\
  \citenamefont {Bahl}}]{peterson2018}%
  \BibitemOpen
  \bibfield  {author} {\bibinfo {author} {\bibfnamefont {C.~W.}\ \bibnamefont
  {Peterson}}, \bibinfo {author} {\bibfnamefont {W.~A.}\ \bibnamefont
  {Benalcazar}}, \bibinfo {author} {\bibfnamefont {T.~L.}\ \bibnamefont
  {Hughes}},\ and\ \bibinfo {author} {\bibfnamefont {G.}~\bibnamefont {Bahl}},\
  }\bibfield  {title} {\bibinfo {title} {A quantized microwave quadrupole
  insulator with topologically protected corner states},\ }\href
  {https://doi.org/10.1038/nature25777} {\bibfield  {journal} {\bibinfo
  {journal} {Nature}\ }\textbf {\bibinfo {volume} {555}},\ \bibinfo {pages}
  {346} (\bibinfo {year} {2018})}\BibitemShut {NoStop}%
\bibitem [{\citenamefont {Noh}\ \emph {et~al.}(2018)\citenamefont {Noh},
  \citenamefont {Benalcazar}, \citenamefont {Huang}, \citenamefont {Collins},
  \citenamefont {Chen}, \citenamefont {Hughes},\ and\ \citenamefont
  {Rechtsman}}]{noh2018}%
  \BibitemOpen
  \bibfield  {author} {\bibinfo {author} {\bibfnamefont {J.}~\bibnamefont
  {Noh}}, \bibinfo {author} {\bibfnamefont {W.~A.}\ \bibnamefont {Benalcazar}},
  \bibinfo {author} {\bibfnamefont {S.}~\bibnamefont {Huang}}, \bibinfo
  {author} {\bibfnamefont {M.~J.}\ \bibnamefont {Collins}}, \bibinfo {author}
  {\bibfnamefont {K.~P.}\ \bibnamefont {Chen}}, \bibinfo {author}
  {\bibfnamefont {T.~L.}\ \bibnamefont {Hughes}},\ and\ \bibinfo {author}
  {\bibfnamefont {M.~C.}\ \bibnamefont {Rechtsman}},\ }\bibfield  {title}
  {\bibinfo {title} {Topological protection of photonic mid-gap defect modes},\
  }\href {https://doi.org/10.1038/s41566-018-0179-3} {\bibfield  {journal}
  {\bibinfo  {journal} {Nat. Photon.}\ }\textbf {\bibinfo {volume} {12}},\
  \bibinfo {pages} {408} (\bibinfo {year} {2018})}\BibitemShut {NoStop}%
\bibitem [{\citenamefont {Imhof}\ \emph {et~al.}(2018)\citenamefont {Imhof},
  \citenamefont {Berger}, \citenamefont {Bayer}, \citenamefont {Brehm},
  \citenamefont {Molenkamp}, \citenamefont {Kiessling}, \citenamefont
  {Schindler}, \citenamefont {Lee}, \citenamefont {Greiter}, \citenamefont
  {Neupert} \emph {et~al.}}]{imhof2018}%
  \BibitemOpen
  \bibfield  {author} {\bibinfo {author} {\bibfnamefont {S.}~\bibnamefont
  {Imhof}}, \bibinfo {author} {\bibfnamefont {C.}~\bibnamefont {Berger}},
  \bibinfo {author} {\bibfnamefont {F.}~\bibnamefont {Bayer}}, \bibinfo
  {author} {\bibfnamefont {J.}~\bibnamefont {Brehm}}, \bibinfo {author}
  {\bibfnamefont {L.~W.}\ \bibnamefont {Molenkamp}}, \bibinfo {author}
  {\bibfnamefont {T.}~\bibnamefont {Kiessling}}, \bibinfo {author}
  {\bibfnamefont {F.}~\bibnamefont {Schindler}}, \bibinfo {author}
  {\bibfnamefont {C.~H.}\ \bibnamefont {Lee}}, \bibinfo {author} {\bibfnamefont
  {M.}~\bibnamefont {Greiter}}, \bibinfo {author} {\bibfnamefont
  {T.}~\bibnamefont {Neupert}}, \emph {et~al.},\ }\bibfield  {title} {\bibinfo
  {title} {Topolectrical-circuit realization of topological corner modes},\
  }\href {https://doi.org/10.1038/s41567-018-0246-1} {\bibfield  {journal}
  {\bibinfo  {journal} {Nat. Phys.}\ }\textbf {\bibinfo {volume} {14}},\
  \bibinfo {pages} {925} (\bibinfo {year} {2018})}\BibitemShut {NoStop}%
\end{thebibliography}
\end{document}